\documentclass[prd,showkeys,showpacs]{revtex4}
\usepackage{graphicx}

\def\be{\begin{equation}}
\def\ee{\end{equation}}
\def\bea{\begin{eqnarray}}
\def\eea{\end{eqnarray}}
\def\Bphi{\mbox{\boldmath $\Phi$}}

\def\hphi{\mbox{\boldmath $\hat\Phi$}}
\def\Bx{\mbox{\boldmath $x$}}

\def\Bal{\mbox{\boldmath $\alpha$}}
\def\Bet{\mbox{\boldmath $\beta$}}
\def\Te{{\cal T}}

\def\dt{\partial_\tau}
\def\dz{\partial_\eta}

\begin{document}
\title{The topological approach to baryon-antibaryon and meson production in rapidly expanding Bjorken rods.}
\author{G. Holzwarth\footnote{e-mail: holzwarth@physik.uni-siegen.de}}
\address{Siegen University, 57068 Siegen, Germany} 
\vspace{3cm}
\begin{abstract}\noindent

The topological approach to baryon-antibaryon production in the chiral phase transition
is numerically simulated for rapidly expanding hadronic systems. For that purpose the
dynamics of the effective chiral field is implemented on a space - rapidity 
lattice. The essential features of evolutions from initial 'hot' configurations into 
final ensembles of (anti-)baryons embedded in the chiral condensate are studied in proper
time of comoving frames. Typical times for onset and completion of the roll-down and  
exponents for the growth of correlations are extracted. Meson and baryon-antibaryon yields 
are estimated. For standard assumptions about initial coherence lengths they are compatible 
with experimental results.  
\end{abstract} 

  
\maketitle

\section{Introduction}

The topological approach to baryon structure and dynamics in the framework of an effective 
action for mesonic chiral fields has achieved a number of
remarkable successes.  The soliton concept~\cite{Skyrme} for baryons provides an impressive
account of spectrum and properties of baryon resonances (essentially without numerous 
'missing resonances')~\cite{Schwesinger}, with predictive power that recently has even led to
the first indications for pentaquarks~\cite{Nakano}. Model-independent
relations between T-matrix elements for meson-baryon scattering~\cite{Hayashi} and explicit results for specific 
channels are well supported by experimental data~\cite{Mattis}. 
The matrix element of the axial singlet current related to the spin content of the proton is naturally 
of the observed order of magnitude~\cite{Brodsky}.
The 'unexpected' behaviour recently found~\cite{Jones} in the ratio of electric and magnetic
proton form factors has been predicted in this approach long ago~\cite{Ho96}. 
The underlying chiral effective action is profoundly based on the
$1/N_c$-expansion of QCD~\cite{Witten}, preserving all  relevant symmetries. Efforts to
include next-to-leading order quantum corrections have brought substantial improvement 
as expected~\cite{Mouss}.  

The manifestations of a chiral phase transition pose another natural challenge for an effectice
theory with a ground state that is characterized by spontaneously broken symmetry.  
The possible formation of disoriented domains~\cite{Bjorken} during the growth of the chiral condensate
has been in the focus of interest for some time. But signatures in terms of anomalous multiplicity ratios 
for differently charged pions have not been observed~\cite{Bearden}, in accordance with theoretical 
conclusions~\cite{Gavin,HoKl02}. Anomalies in anti-baryon production were very early recognized
as possible signals for interesting dynamics~\cite{DeGrand} in that phase transition, and the concept 
to consider baryons as topological solitons in a chiral condensate should lead to quite definite 
expectations for this process. 

Meanwhile, in relativistic heavy-ion collisions at RHIC, very high energy densities are being 
produced in extended spatial regions which are essentially baryon free and well separated in rapidity from the 
nuclear slabs receding from the collision volume. 
The experimental values found in the central rapidity region for the ratio of the integrated
${\bar p}$ to $\pi^-$ yields 
lies between 0.065 and 0.075 \cite{phenix}. This is still too close
to the thermal equilibrium ${\bar p}/\pi^-$ ratio (for a typical plasma temperature of $\Te \sim 200$ MeV),
\be
{\bar p}/\pi^- \sim 2 \exp((m_\pi-m_p)/\Te)=0.035
\ee
 to constitute a clear indication 
for interesting underlying physics. Still, although the experimental result does not look very 
exciting, it still poses a constraint for the possible validity of the soliton concept,
because any conceivable dynamical production process must be able to produce a comparable number.

In the topological approach the number of baryon-antibaryon pairs produced during the 
chiral phase transition depends on two factors: the first is the modulus $|\rho|$ of the average winding density
present in the initial 'hot' field configuration. In analogy to applications in cosmology~\cite{Kibble} and
condensed matter systems~\cite{Zurek} this quantity is closely related to the coherence length for the local
orientations of the chiral field $\Bphi$. Without detailed knowledge about the initial field configurations
this coherence length enters as a parameter and takes away stringent predictive power from the approach.
However, different conjectures about the nature of the initial field ensemble suggest typical ranges for the
coherence lengths which then may be discriminated by the experimentally observed abundancies.    

The second factor is the reduction of the initially present total $n_i=\int |\rho| dV$ through the dynamical 
ordering process, which finally leads to the formation of stable soliton structures embedded in the topologically
trivial ordered chiral condensate of the 'cold' system. The solitons or antisolitons evolve from
topological obstacles which are met by the aligning field orientations, and  develop into their
stable 'cold' form during the course of the evolution. At the end, the same integral $n_f=\int |\rho| dV$
counts the number of finally surviving nontrivial separate structures, so it is identified with the number of
baryons and antibaryons created in the process. The decrease of $n$ during the roll-down 
is reasonably well represented by
a power law $(\tau/\tau_0)^{-\gamma}$ and the exponent $\gamma$ can be measured in numerical simulations.
Evidently, the initial time $\tau_0$ which marks the onset of the evolution, enters here as a second
parameter which further reduces the predictive power of the approach. Fortunately, it turns out that $\gamma$
is rather small, so the dependence on $\tau_0$ is only weak. 

Measuring $\gamma$ and the time $\tau_f$ when the roll-down is completed, presents a typical task for
numerical simulations once the equation of motion (EOM) which governs the field evolutions is implemented
on a lattice. The underlying effectice chiral action is known from other applications, so no additional
parameters enter at this point. In condensed matter applications, a phase transition is generally driven
by an externally imposed quench, or by a dissipative term included in the EOM. In cosmology or in our
present heavy-ion application it is the rapid expansion of the hot volume which drives the cooling process. 
This expansion is efficiently implemented~\cite{Huang} by transforming to rapidity -
proper-time coordinates, i.e. by boosting to the local comoving frame. This is especially convenient if we consider
a system that expands only in one (longitudinal) direction with its transverse scales unchanged, (the Bjorken rod).
The resulting dilution of the longitudinal gradients drives the system towards its global minimum. However, as
there is no genuine dissipation in the system, the total energy approaches a constant which resides in the 
chiral fluctuations around the global minimum. Thus,  the simulations also allow to estimate pion- or sigma-
meson abundancies.

Naturally, before the field configurations can roll down towards the global minimum, the potential 
$V(\Phi^2,\Te)$ which underlies
the EOM must have changed from the 'hot' chirally symmetric form to its 'cold' symmetry-violating form.
But, during the early stages, the evolutions are dominated by local
aligning of the field orientation $\hphi$. During this phase the form of the potential is 
not important. So its time dependence can be replaced by a sudden quench where the 'hot' field 
configuration is exposed to the 'cold' potential $V(\Phi^2,\Te=0)$, from the outset at initial time $\tau_0$ . 
In the following, for definiteness we make use of this sudden quench approximation, (although the simulations,
of course, allow to study other cases as well).   

For the sake of simplicity we first discuss all relevant features for the case of the 2-dimensional 
$O(3)$-model, with only one spatial dimension transverse to the longitudinal rapidity coordinate.   
Except for computational complexity the extension to the 3-dimensional $O(4)$-field presents no essential
new features. The effective action, its transformation to the Bjorken frame, and the resulting EOM
are presented in section II. It is important for the choice of the initial ensemble of field configurations
that it allows in a convenient way to monitor the initial coherence lengths because they are the crucial
parameters for the final baryon-antibaryon multiplicities. We choose an isotropic Gaussian random ensemble of 
field fluctuations in momentum space which is characterized by a temperature-like parameter to be able to 
compare with other approaches. Of course, this is not necessary. In fact, even at initial time $\tau_0$
the longitudinally expanding
Bjorken rod need not be an isotropic system, and it may be physically justified to distinguish already in the
initial ensemble two different, longitudinal and transverse correlation lengths. This is easy to incorporate, 
but in section III we present initial conditions which are locally isotropic.

As discussed elsewhere~\cite{Ho03} stable solitons shrink in a spatially expanding frame. Therefore, lattice
implementations of their dynamics will necessarily involve lattice artifacts after some time. These are 
discussed in section IV. They can be isolated and subtracted from the physically interesting quantities.

In section V the essential features of typical evolutions are discussed. Estimates for the times of onset
and completion of the roll-down are obtained, and the dynamical exponents for the growth of correlation lengths and
decrease of defect number are established and compared. The spectrum of the fluctuations remaining after the
roll-down is considered and finally the mesonic and baryonic multiplicities are obtained.

The extension to the physically interesting 3+1-dimensional $O(4)$-field is discussed in section VI. The 
topological generalization is well known, the additional transverse dimension is of little influence for the 
growth exponents. However, the coupling constants in the effective action here are related to physical
quantities, so they are known (except for some uncertainty concerning the $\sigma$-mass), and the results can
be compared with experimentally determined abundance ratios.

Of course, it would be desirable to obtain a very definite answer whether the topological approach to
antibaryon production in a chiral phase transition is validated or ruled out by the data. However, with our poor
knowledge about the initial conditions in the hot plasma after a heavy-ion collision, we cannot expect 
much more than allowed ranges for the relevant parameters, which hopefully overlap with standard ideas
about coherence lengths and formation times.

\section{The effective action in the Bjorken frame.}

For simplicity we first discuss the 2+1 dimensional $O(3)$ model. It is defined in terms of the dimensionless 
$3$-component field $\Bphi = \Phi \hphi$ with unit-vector field $\hphi$, ($\hphi
\cdot \hphi =1$), and modulus ('bag'-)field $\Phi$,
with the following lagrangian density in $2+1$ dimensions $(x,z,t)$

\be\label{lag}
{\cal{L}} = f_\pi^2\left({\cal{L}}^{(2)}+{\cal{L}}^{(4)}+{\cal{L}}^{(0)}\right)
\ee
($f_\pi^2$ is an overall constant of dimension [mass$^1$], so the physical fields $f_\pi\Bphi$ are of 
mass-dimension [mass$^{1/2}$]).
The second-order part ${\cal{L}}^{(2)}$ comprises the kinetic terms of the linear $\sigma$ model 

\be \label{L2}
{\cal{L}}^{(2)}=\frac{1}{2} \partial_\mu \Bphi \partial^\mu \Bphi,
\ee
${\cal{L}}^{(4)}$ is the four-derivative 'Skyrme'-term (which involves only the unit-vector field $\hphi$)
defined in terms of the topological current $\rho_\mu$ 
\be\label{top}
\rho^\mu = \frac{1}{8 \pi} \epsilon^{\mu \nu \rho} \hphi \cdot
( \partial_\nu \hphi \times \partial_\rho \hphi ) ,
\ee
(which satisfies $\partial_\mu \rho^\mu = 0$)
\be \label{L4}
{\cal{L}}^{(4)}=-\lambda\ell^2  \:\varrho_\mu\varrho^\mu=
-\frac{2\lambda\ell^2}{(8\pi)^2}\left[ (\partial_\mu \hphi \partial^\mu
\hphi)^2 -(\partial_\mu \hphi \partial_\nu \hphi)(\partial^\mu \hphi\partial^\nu \hphi) \right], 
\ee
and ${\cal{L}}^{(0)}$ contains the $\Phi^4$ potential and an explicit symmetry-breaker in 3-direction
\be
\label{pot}
{\cal{L}}^{(0)}=-V(\Bphi,\Te)=-\frac{1}{\ell^2}\left( \frac{\lambda}{4}\left( \Bphi^2-f(\Te)^2 \right)^2
- H \Phi_3 \right)  -\mbox{const.}
\ee
with dimensionless coupling constants $\lambda$ and $H$, and
\be\label{f2}
f^2(\Te)= f_0^2(\Te)-\frac{H}{\lambda f_0(\Te)}.
\ee
 This choice ensures that the global 
minimum of the potential $V(\Bphi,\Te)$ is always located at $\Bphi_0=(0,0,f_0(\Te))$. Generically, the function 
$f_0^2(\Te)$ decreases from $f_0^2=1$ at $\Te=0$ towards zero for large $\Te$.  
The constant in the potential (\ref{pot})
is chosen such that the value of the potential $V$ at $\Bphi=0$ is independent of $\Te$, (given by the constant
$V(0,\Te)=(\lambda +2H)/(4\ell^2)$), 
and at the $(\Te=0)$-minimum $\Bphi=\Bphi_0=(0,0,1)$ we have $V(\Bphi_0,\Te=0)=0$. 

The masses of the $\pi$- and $\sigma$-fluctuations $(\pi_1,\pi_2, f_0+\sigma)$ around this minimum  are
\be\label{masses}
m_\pi^2=\frac{H}{\ell^2 f_0},~~~~~~~~~~~~m_\sigma^2=\frac{2\lambda f_0^2}{\ell^2} + m_\pi^2.
\ee 
Without explicit symmetry breaking, $H=0$, we assume that $f^2(\Te)$ changes sign at $\Te=\Te_c$, such that
$\Bphi_0=(0,0,0)$  and $m_\sigma^2=m_\pi^2=m^2=\lambda |f^2| / \ell^2$ for $\Te>\Te_c$. 

The parameter $\ell$ (with dimension of a length) which we have separated out from the coupling constants of
potential and Skyrme term
can be absorbed into the spatial coordinates $\Bx$. So it characterizes the spatial radius 
of stable extended solutions (which scales like $1/\sqrt{f^2}$). As $\ell$ simply sets the spatial scale,
it could be put equal to one, as long as no other (physical or artificial) length  scales are relevant. For
lattice implementations, however, the lattice constant $a$ and the size of the lattice $(Na)$
set (usually unphysical) scales. 
To avoid artificial scaling violations we have to ensure that the size of physical structures (like solitons) 
is large as compared to the lattice constant $a$ and small as compared to the lattice size $Na$. So,
for numerical simulations we have to choose $1\ll \ell/a \ll N$. 
It has been shown in ref.\cite{HoKl01} that for solitons which extend over more than at least 4-5 lattice units the 
energy $E_B$ is independent of $\ell/a$.
So, in the following we will adopt $\ell/a\sim 5$ as sufficiently large. This appears also as 
physically reasonable, if we consider
typical lattice constants of 0.2 fm and baryon radii of about 1 fm. On the other hand this will require lattice
sizes of at least $N \sim 50$ to avoid boundary effects for the structure of individual solitons.
Unfortunately, in the Bjorken frame which we shall use in the following, the longitudinal extension of
stable solitons shrinks like $\tau^{-1}$ as function of proper time $\tau$. 
This means that after times of order $\ell$ the simulations will be
influenced by lattice artifacts, which may even dominate for large times.

For rapid expansion in (longitudinal) $z$-direction we perform the transformation from $(z,t)$ to
locally comoving frames ($\eta$,$\tau$) with proper time $\tau$ and rapidity $\eta$, defined through
\begin{eqnarray}\label{trafo}
t&=&\tau \cosh\eta,~~~~~~~~~~\tau~=~\sqrt{t^2-z^2},~~~~~~~~~~~~~
\partial_t~=~\cosh\eta\; \dt -\frac{\sinh\eta}{\tau}\dz \nonumber\\
z&=&\tau \sinh\eta,~~~~~~~~~~\eta~=~\mbox{atanh}\left(\frac{z}{t}\right),~~~~~~~~~~~~
\partial_z~=~-\sinh\eta\; \dt +\frac{\cosh\eta}{\tau}\dz.
\end{eqnarray}
Inserting (\ref{trafo}) into (\ref{L2}) and (\ref{L4}) 
leaves the form of ${\cal{L}}^{(2)}$ and ${\cal{L}}^{(4)}$ invariant, with $\partial_t$ replaced by
$\dt$, and $\partial_z$ replaced by $\frac{1}{\tau}\dz$. 
The specific structure of the Skyrme term again eliminates all terms with four $\tau$- or $\eta$-derivatives. 
For the effective action we take the integration boundaries from $-\infty$ to $+\infty$ 
for rapidity $\eta$ and for the transverse coordinate $x$.
The 3-dimensional space-time volume element $dx\:dz\:dt$ is replaced by $\tau \:dx \:d\eta \:d\tau$. 
Therefore, in a separation
of the action ${\cal S}$ in kinetic terms $T$, gradient terms $L$, and the potential $U$, 
\be \label{S}
{\cal S}=\int d\tau \int_{-\infty}^{+\infty}\!{\cal L} \;d\eta \:dx  
=\int(T_{\bot}+T_{\|}-L_{\bot}-L_{\|}-U) d\tau
\ee
the longitudinal $\|$-terms involving rapidity gradients carry a factor $1/\tau$, while all other 
terms carry a factor $\tau$. 
So we have
\bea
T_{\bot}&=&\tau \int \left\{\frac{1}{2} (\dt\Bphi \dt\Bphi)
+\frac{\lambda\ell^2}{(4\pi)^2}\left[ \frac{\Bphi}{\Phi^3}\cdot(\dt \Bphi\times
\partial_x \Bphi) \right]^2\right\}\;d\eta \:dx,\label{int1} \\
T_{\|}&=&\frac{1}{\tau} \int \left\{
\frac{\lambda\ell^2}{(4\pi)^2}\left[ \frac{\Bphi}{\Phi^3}\cdot(\dt \Bphi\times
\dz \Bphi)\right]^2\right\}\;d\eta \:dx,\label{int2}\\
L_{\bot}&=&\tau \int \left\{\frac{1}{2} (\partial_x\Bphi \partial_x\Bphi)
\right\}\;d\eta \:dx,\label{int3}\\
L_{\|}&=&\frac{1}{\tau} \int \left\{\frac{1}{2} (\dz\Bphi \dz\Bphi)
+\frac{\lambda\ell^2}{(4\pi)^2}\left[ \frac{\Bphi}{\Phi^3}\cdot(\dz \Bphi\times
\partial_x \Bphi) \right]^2\right\}\;d\eta \:dx, \label{int4}\\
U&=&\tau \int \left\{\frac{\lambda}{4\ell^2}\left( \Bphi^2-f^2 \right)^2
- \frac{H}{\ell^2} \Phi_3 +\mbox{const.}\right\}\;d\eta \:dx.  \label{int5}
\eea
Variation of ${\cal S}$ with respect to $\Bphi$ leads to the equation of motion (EOM).

The contributions of ${\cal{L}}^{(4)}$ to the longitudinal and transverse parts $T_{\|}$ and $T_{\bot}$
of the kinetic energy cause certain numerical difficulties for the implementation of the EOM on a lattice.
They require at every timestep the inversion of matrices which depend on gradients of the unit-vectors $\hphi$,
which multiply first and second time-derivatives of the chiral field. This can be troublesome in areas
where the unit-vectors are aligned, and can be poorly defined in regions where the unit-vectors vary
almost randomly for next-neighbour lattice points (i.e. for initially random configurations, or near the center
of defects). In any case, stabilizing the evolutions requires extremely small timesteps and leads to
very time-consuming procedures. Although these problems can be handled, we have compared the results
with evolutions where the kinetic energy is taken from ${\cal{L}}^{(2)}$ alone.
For coupling strenghths $\lambda\ell^2$ within reasonable limits,
we find that the resulting differences do not justify the large additional expense caused by the fourth-order
kinetic contributions. Evidently, the reason is, that the EOM determines the field-velocities (depending on
the functional form of the kinetic energy) in such a way that the numerical value of the total kinetic energy
is not very sensitive to its functional form. We therefore use in the following an effective action where
the kinetic terms (\ref{int1}) and (\ref{int2}) are replaced by
\be\label{Tkin}
T_{\bot}=\frac{\tau}{2}\int(\dt\Bphi \dt\Bphi)\,d\eta \,dx,~~~~~~~~~~~~~~~T_{\|}=0.
\ee 
With this simplification the EOM is 
\be \label{EOM}
\frac{1}{\tau}\partial_\tau\Bphi+\partial_{\tau\tau}\Bphi-\partial_{xx}\Bphi-
\frac{1}{\tau^2}\partial_{\eta\eta}\Bphi+\frac{\lambda}{\ell^2}(\Phi^2-f^2)\Bphi
-\frac{H}{\ell^2} \hat{\mbox{\boldmath $e$}}_3 +\frac{\lambda\ell^2}{\tau^2}\frac{\delta \rho_0^2}{\delta\Bphi}=0.
\ee
This form has the big advantage that we can make use of the geometrical meaning of the winding density
$\rho_0$ as the area of a spherical triangle, bounded by three geodesics on a 2-dimensional spherical surface.
In closed form it is expressed through the unit-vectors pointing to its corners, and does not involve gradients.
So this allows for a very accurate and fast lattice implementation of the last term in the EOM.

\section{Initial configurations} 

We assume that at an initial proper time $\tau_0$ the system consists of a hadronic fireball 
with energy density $\varepsilon_0$ stored in a random ensemble
of hadronic field fluctuations. Subsequently, for $\tau > \tau_0$, it is subject to EOM (\ref{EOM}). 
The initial condition and the symmetry of the action imply
boost invariance, i.e. the system looks the same in all locally comoving frames, so it is 
sufficient to consider its dynamics in a rapidity slice of size $\Delta\eta$ near midrapidity $\eta=0$,
which constitutes a section of the initially created Bjorken rod with transverse extension 
${\cal A}$. The energy $E=T+L+U$ in this slice then is given by an  $\eta$-integral 
which extends over the finite rapidity interval $\Delta\eta$ and represents the energy 
contained in a comoving volume ${\cal V}= \tau \Delta\eta {\cal A}$.
Due to the symmetry of the initial condition this comoving volume grows with increasing proper time $\tau$ into
spatial regions with high energy density, therefore $E$ contains contributions which increase with $\tau$. 
The average energy density $\varepsilon=E/{\cal V}$ satisfies $d\varepsilon/d\tau \leq 0$.

For numerical simulations we implement the configurations $\Bphi(x,\eta,\tau)$ on a rectangular 
lattice $(x,\eta) = (ia,jb)$ ($i,j=1...N$) with lattice constants $a$ for the transverse coordinate and $b$ for
the rapidity lattice.
We define the initial configurations $\Bphi_{ij}$  at the lattice sites ($i,j$) as Fourier transforms
of configurations $\tilde {\Bphi}_{kl}$ on a momentum lattice 
\be  \label{Fourier}
\Bphi_{ij}=\frac{1}{N}\sum_{k,l=-N/2+1}^{N/2} 
\frac{1}{2}\left( e\:^{\mbox{\small i}\frac{2\pi}{N}(i\cdot k +j\cdot l)} 
\tilde{\Bphi}_{kl}+ c.c.\right),
\ee
with $\tilde{\Bphi}^*_{kl}=\tilde{\Bphi}_{-k-l}$. Inversely, the real parts $\Bal_{kl}$ and the imaginary parts 
$\Bet_{kl}$ of $\tilde {\Bphi}_{kl}$ are obtained from the real configuration
$\Bphi_{ij}$ through
\bea \label{albet}
\Bal_{kl}& =&\frac{1}{N}\sum_{i,j=1}^{N}\cos\frac{2\pi}{N}(i k +j l)\;\Bphi_{ij}=\Bal_{-k-l}\\
\Bet_{kl} &=&-\frac{1}{N}\sum_{i,j=1}^{N}\sin\frac{2\pi}{N}(i k +j l)\;\Bphi_{ij}=-\Bet_{-k-l},
\eea
so we obtain the spectral power $P_{pq}$ of the configurations (or a specific component of it) 
at any time $\tau$ from 
\be
P_{pq}=\tilde {\Bphi}_{kl}\cdot \tilde {\Bphi}_{kl}^*=\Bal_{kl}\cdot\Bal_{kl}+\Bet_{kl}\cdot\Bet_{kl}
\ee
for any transverse or longitudinal momentum $(p,q)= \frac{2\pi}{aN}(k,l)$, 
for $(k,l = -N/2+1,..., N/2)$. 

For the initial configurations at $\tau=\tau_0$ the real and imaginary parts of each of the three 
components of $\tilde {\Bphi}_{kl}$  
at each momentum-lattice point ($p,q$) are chosen 
randomly from a Gaussian deviate $G_{kl}(\tilde{\Phi})$ with $kl$-dependent width $\sigma_{kl}$,
\be \label{Gauss}
G_{kl}(\tilde{\Phi})=\frac{1}{\sqrt{2\pi \sigma_{kl}^2}} \exp\left(-\frac{\tilde\Phi^2}{2\sigma_{kl}^2}\right),
~~~~\mbox{with}~~~~~
\sigma_{kl}^2=\frac{\sigma_0^2}{Z}\exp\left(-\frac{\sqrt {p^2+q^2+m^2}}\Te\right),
\ee
with normalization $Z$ chosen in such a way that
\be\label{modsum}
\sum_{k,l=-N/2+1}^{N/2}\sigma_{kl}^2 = N^2\sigma_0^2.
\ee
(In the continuum limit $(a\rightarrow 0$, $N\rightarrow\infty)$ we have 
$Z=\frac{\Te^2}{2\pi}\left(1+\frac{m}{\Te}\right)e^{-m/\Te}.)$

In other words, we choose a Boltzmann distribution for the average occupation numbers 
$n_{kl} =\langle\langle \tilde\Phi_{kl}\tilde\Phi_{kl}^*\rangle\rangle
 = \sigma_{kl}^2$ for each field component, as for relativistic (non-interacting)
particles with mass $m$. Here the mass $m^2$ is defined by the absolute value 
\be\label{mass}
m^2(\Te)=\frac{\lambda}{\ell^2} |f^2(\Te)|
\ee 
for the fluctuations around $\Bphi=0$ in the symmetric potential (\ref{pot}) at the initially high
temperature $\Te=\Te_0$, where $f^2(\Te)$ is negative. 
The amplitude $\sigma_0^2$ plays the role of a fugacity
\be
\sigma_0^2=\exp(-\mu/\Te)
\ee
for negative chemical potential $\mu$. In the temperature range which we consider $(0.05<a\Te<0.8)$
(cf. fig.(\ref{fig3})) a suitable value for $\mu$ is $a\mu\sim -0.6$. (With this choice the 
average amplitude of the chiral field is not subject to abrupt deviations from its initial value 
immediately after the onset of the dynamical evolution).  

We assume isotropy of the initial ensemble with respect to rotations in $O(3)$-space such that
the three components of the field fluctuations $\tilde\Phi^\alpha_{kl}$ ($\alpha=1,2,3$)
have the same average square amplitude $\sigma_{kl}^2$. By picking each component independently
at each point $(k,l)$ from the Gaussian ensemble, different components are uncorrelated and equal components 
at different points (on the momentum lattice) are also uncorrelated,   
\bea
\langle\langle \tilde\Phi^\alpha_{kl} \tilde\Phi^{\beta *}_{k'l'} \rangle\rangle &=&
\langle\langle \alpha^\alpha_{kl} \alpha^\beta_{k'l'} \rangle\rangle +
\langle\langle \beta^\alpha_{kl} \beta^\beta_{k'l'} \rangle\rangle \nonumber \\
&=&\sigma_{kl}^2 \delta_{\alpha\beta}
\left( \frac{1}{2}(\delta_{k k'}\delta_{l l'} +\delta_{-k k'}\delta_{-l l'})
+\frac{1}{2}(\delta_{k k'}\delta_{l l'} -\delta_{-k k'}\delta_{-l l'})\right) 
=\sigma_{kl}^2 \delta_{\alpha\beta}\delta_{k k'}\delta_{l l'}.
\eea
Together with (\ref{Fourier}) this leads to the fluctuation in the real field configurations
\be \label{fluc}
\langle\langle \Phi^\alpha_{ij}\Phi^{\beta }_{ij} \rangle\rangle 
=\delta_{\alpha\beta}\frac{1}{N^2}\sum_{k,l=-N/2+1}^{N/2} \!\!\! \sigma_{kl}^2
= \delta_{\alpha\beta} \; \sigma_0^2
\ee
which is, of course, independent of the lattice point $(i,j)$. Its magnitude is controlled by the constant
$\sigma_0^2$ in (\ref{Gauss}). It should be sufficiently small to keep the amplitudes of the average
initial fluctuations small. 
On the lattice the upper limit for the momenta $p$,$q$ is $\frac{\pi}{a}$, (i.e. $k,l=N/2$). So, as long as 
\be \label{LamT}
\Te \ll \frac{\pi}{a},
\ee
the lattice cut-off (upper limit momentum) imposed by the finite lattice constant 
is unimportant because the corresponding states are almost unoccupied. 
Note that periodicity and antisymmetry of the imaginary parts  in (\ref{albet}) requires
that  $\Bet_{kl}$ vanishes if both $k$ and $l$ are multiples of $N/2$. With the condition
(\ref{LamT}) satisfied, this holds with good accuracy also for the initial configuration picked 
randomly from the ensemble (\ref{Gauss}). 
 
 The average number of topological defects in a random ensemble of vector configurations
 is closely related to the characteristic angular coherence  length  in that ensemble.
 Therefore, it will be necessary to measure the (equal-time) correlation functions
 for the unit-vector fields $\hphi$ for the evolving ensembles.
 In order to have an analytical result at least for the initial configurations
 (where length and orientation of the 3-vectors are uncorrelated), it is easier to consider
 the correlations among the full vectors $\Bphi$.
Therefore, we define normalized transverse and longitudinal correlation functions
 \bea \label{corrfct}
 C_\bot(i)&=&\frac{1}{3\sigma_0^2N^2}
 \left[\langle\langle\sum_{m,n=1}^{N}\Bphi_{mn}\cdot\Bphi_{m+i,n}\;\rangle\rangle
 -\frac{1}{N^2}\langle\langle\sum_{m,n=1}^{N}\Bphi_{mn}\;\rangle\rangle \cdot 
 \langle\langle\sum_{k,l=1}^{N}\Bphi_{kl}\;\rangle\rangle\right],\nonumber \\
 C_\|(i)&=&\frac{1}{3\sigma_0^2N^2}
 \left[\langle\langle\sum_{m,n=1}^{N}\Bphi_{mn}\cdot\Bphi_{m,n+i}\;\rangle\rangle
 -\frac{1}{N^2}\langle\langle\sum_{m,n=1}^{N}\Bphi_{mn}\;\rangle\rangle \cdot 
 \langle\langle\sum_{k,l=1}^{N}\Bphi_{kl}\;\rangle\rangle\right],
 \eea
 with transverse coherence  lengths $R_\bot$ and longitudinal (dimensionless) coherence  rapidity $R_\|$
 defined through
 \be \label{cut}
 C(i)< \frac{1}{e}~~~~~~~~~~\mbox{for}~~~~ i > \frac{R_\bot}{a},~~~~\mbox { or}~~~ i > \frac{R_\|}{b}
 \ee
 respectively.
 For the initial ensemble (\ref{Gauss}) the correlations
 are, of course, isotropic on the lattice, i.e. 
 \be \label{initcorr}
 \frac{R_\bot}{a}=\frac{R_\|}{b}=\frac{R_0}{a}
 \ee
 with initial spatial coherence  length $R_0$.
 In the continuum limit $(a\rightarrow 0$, $N\rightarrow\infty)$, we obtain
 $C(r)$ as function of the spatial distance $r$ (or rapidity $\eta=r (b/a)$)
 \be  \label{corrinit}
 C(r)=
 \frac{e^{-\frac{m}{\Te}\left(\sqrt{1+r^2\Te^2}-1\right)}}{\left(1+r^2\Te^2\right)^{3/2}}
 \left(\frac{1+\frac{m}{\Te}\sqrt{1+r^2\Te^2}}{1+\frac{m}{\Te}}\right).
 \ee
 Specifically, putting $m=0$, the coherence  length $R$ as defined in (\ref{cut}) is  
 \be \label{Rm0}
 R=\frac{\sqrt{e^{2/3}-1}}{\Te} \approx \frac{0.97}{\Te}. 
 \ee
 This allows to put limits on the range of temperatures which can be reasonably represented
 on the lattice. Typically, for lattice size of $N\sim 100$, $\Te$ should lie within the range  
 from about  0.02 to about 0.8 inverse lattice units. For smaller values the inital coherence  
 length already covers more than half of the lattice so almost no defects will fit on the lattice,
 for larger values the correlation lengths approach the lattice constant.
  It may be noted that with (\ref{mass}), for $(\ell/a)\sim 5$ and $(a\Te)\sim 0.1$, 
  the ratio $m/\Te$ is not very small, so generally we expect appreciable deviations from the 
 $\Te^{-1}$ scaling in (\ref{Rm0}), (e.g. for $(\ell/a)=4$ we find $(R/a)\sim (a\Te)^{-0.8}$, 
 cf. fig.\ref{fig3}).
 
 During the evolution in the Bjorken frame the correlations rapidly become anisotropic. 
 We then conveniently define an average coherence  
 length $\bar R$ through
 \be \label{avcorr}
 \frac{a^2}{{\bar R}^2}=\frac{1}{2}\left( \frac{a^2}{R_\bot^2}+\frac{b^2}{R_\|^2}\right).
 \ee 
This may be compared to the coherence  radius obtained from the angular-averaged correlation function 
\be \label{avcorrfct}
\bar C(r)
=\frac{1}{3\sigma_0^2N^2}
\left[\langle\langle\sum_{m,n=1}^{N}\sum_{i,j}\Bphi_{mn}\cdot\Bphi_{m+i,n+j}\;\rangle\rangle
-\frac{1}{N^2}\langle\langle\sum_{m,n=1}^{N}\Bphi_{mn}\;\rangle\rangle \cdot 
 \langle\langle\sum_{k,l=1}^{N}\Bphi_{kl}\;\rangle\rangle\right]
\ee 
where the $i,j$-sum indicates an average over all lattice points in a narrow circular ring with radius $r$ 
around the lattice point $mn$.

The essential characteristics of the evolutions are not very sensitive to the choice of the 
initial time derivatives. (They can as well be put to zero.) The equations of motion very quickly establish
appropriate velocities. Of course, the absolute value of the total energy depends on that choice.
For the simulations presented in the following we construct in analogy to the initial 
configurations (\ref{Fourier}) an initial ensemble of time derivatives 
through
\be
(\partial_\tau\Bphi)_{ij}=\frac{\mbox{i}}{N}\sum_{k,l=-N/2+1}^{N/2} \frac{\omega_{kl}}{2} 
\left( e\:^{\mbox{\small i}\frac{2\pi}{N}(i\cdot k +j\cdot l)} \tilde{\Bphi}_{kl}- c.c.\right),
~~~~~~\mbox{with}~~~~~
\omega_{kl}=\sqrt{\left(\frac{2\pi}{aN}k\right)^2+\left(\frac{2\pi}{a N}l\right)^2 + m^2}.
\ee 
The Fourier coefficients $\tilde{\Bphi}_{kl} $ again are picked randomly from 
the same Gaussian deviate (\ref{Gauss}).

\section {Shrinking solitons in comoving frames}
Let $\Bphi^{(s)}(x,z)$ be a static soliton solution of the model (\ref{lag}) in its $(x,z)$ rest 
frame, which minimizes the static energy $E=L+U$ with a finite value for the 
soliton energy $E=E_0$. 
After the transformation to the Bjorken frame, the configuration 
$\Bphi_\tau^{(s)}(x,\eta)=\Bphi^{(s)}(x,\tau\eta)$ then describes a static solution of the action in the
comoving $(x,\eta)$ frame at proper time $\tau$ (where $\partial_z$ is replaced by $(1/\tau) \partial_\eta$),
for the same value of $E_0$. It represents a soliton with the same finite radius 
in transverse $x$-direction as before,
but with its radius in longitudinal $\eta$-direction shrinking like $1/\tau$ with increasing proper time $\tau$.
The total energy $E_0$ of this shrinking soliton is, of course, independent of $\tau$.
(Naturally, this consideration strictly applies 
only to the adiabatic case, where $\tau$ is considered as a parameter. In the dynamical ordering process 
the evolution of the solitons towards their static form may appreciably lag behind the actual progress 
of proper time.)
\begin{figure}[h]
\centering
\includegraphics[width=9.5cm,height=13cm,angle=-90]{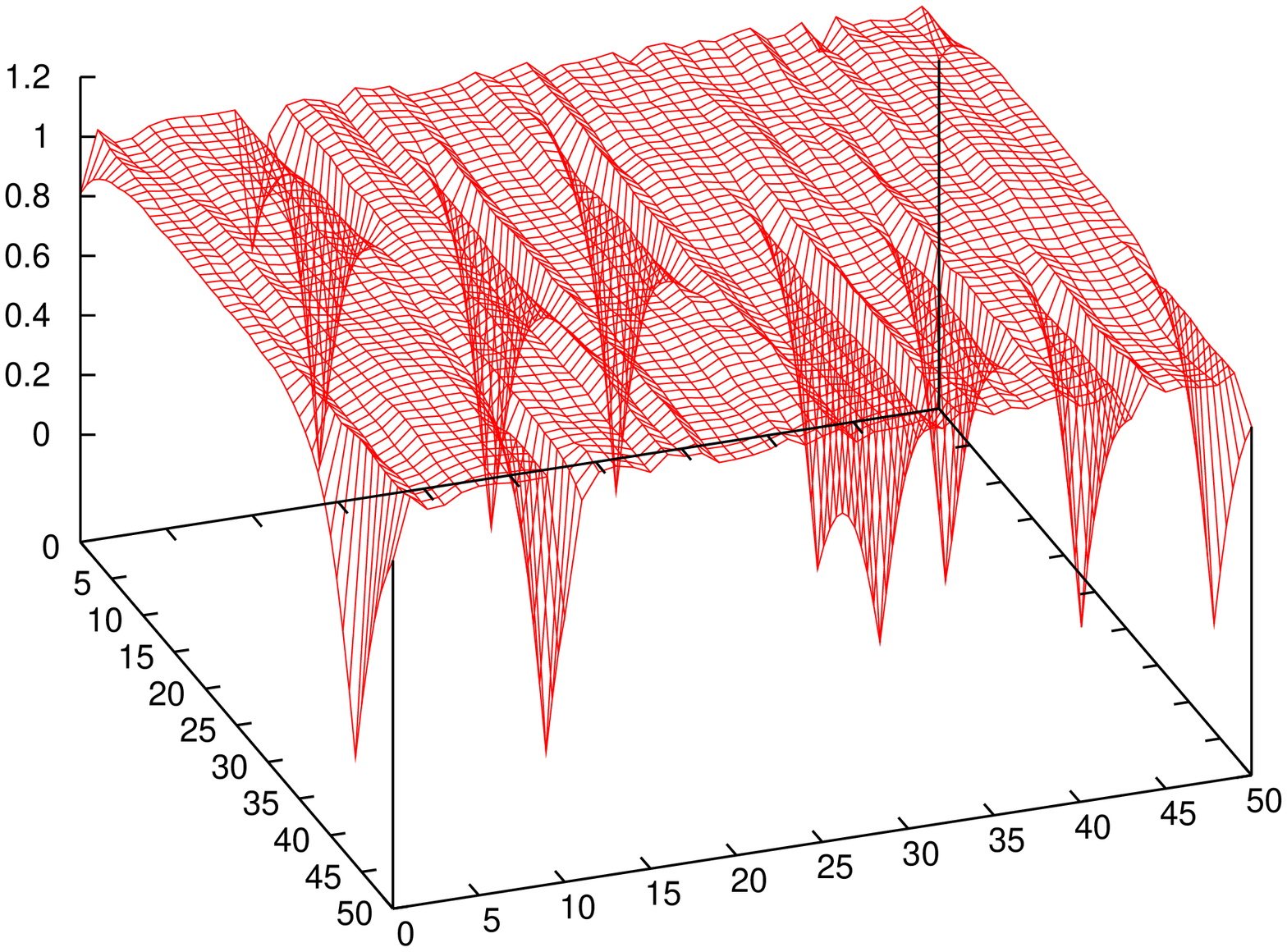}
\includegraphics[width=9.5cm,height=13cm,angle=-90]{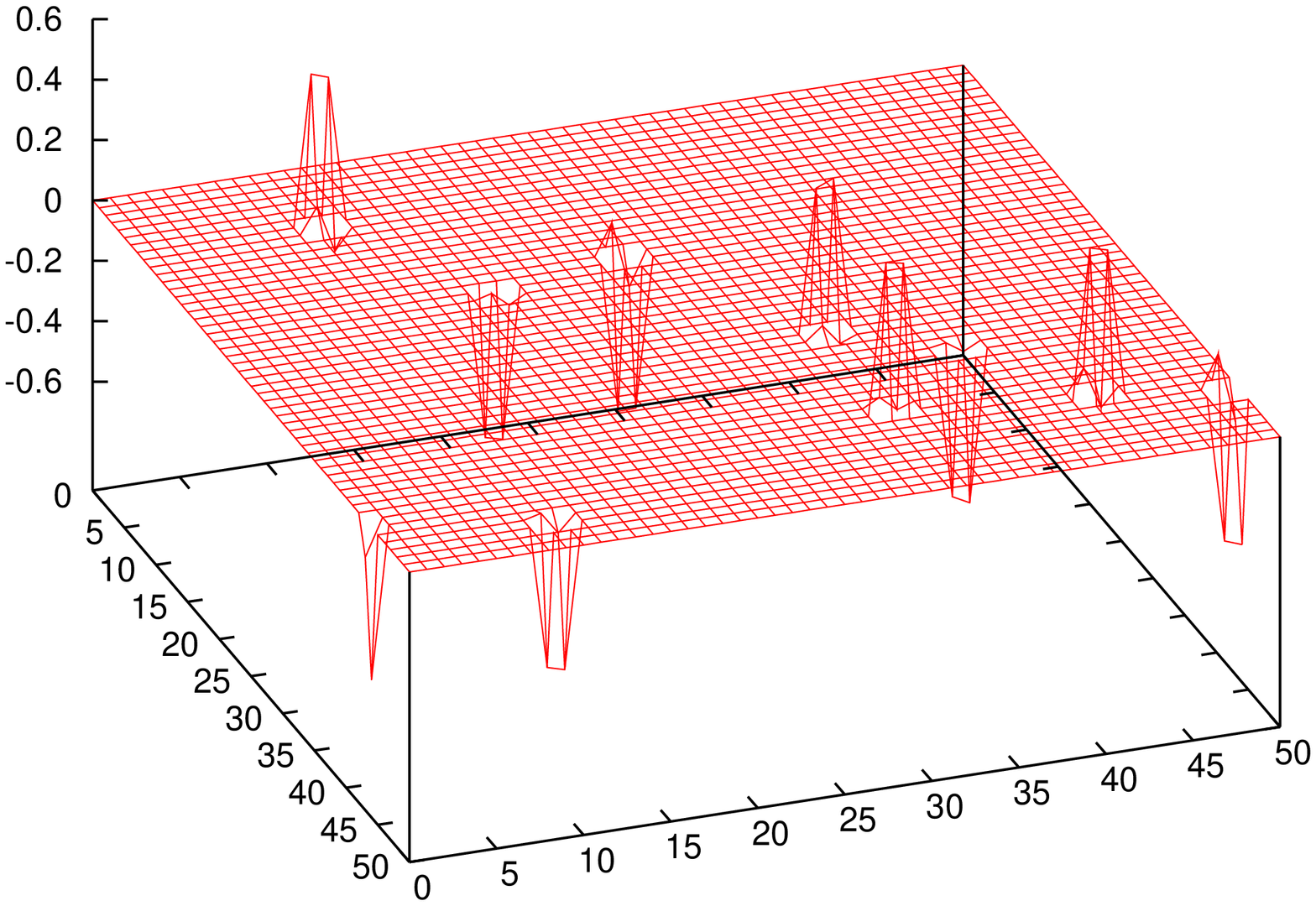}
\caption{Soliton configuration after a typical evolution on a 50$\times$50 lattice (for $\lambda=1, 
\ell/a=4, \sigma_0=0.2, H=0.2, a\Te=0.2$) at time $\tau/\tau_0=1000$, i.e. long after completion
of the roll-down. The bag-field $|\Bphi|$ of the solitons (upper part) is sqeezed longitudinally to 
lattice-unit size; the positive or negative winding densities (lower part) are located at the 
center of the bags.} 
\label{fig1}
\end{figure}
For lattice implementations,
with the typical spatial radius of the stable solitons given by $\ell$,  
the longitudinal extension of the solitons for times $\tau \gg \ell $ has shrunk
down to (dimensionless rapidity-)lattice-unit size and longitudinally adjacent solitons no longer interact. 
In transverse direction, however, the solitons develop their stable size of $\ell/a$ lattice units,
they keep interacting, attracting close neighbours or annihilating with overlapping antisolitons
(cf. fig.\ref{fig1}).

For solitons shrinking longitudinally down to lattice-unit size the energy will begin to deviate 
from the value $E_0$ as soon as the longitudinal extent covers merely a few lattice units.
To get an approximate idea for the energy limit 
let us assume that a single separate soliton finally degenerates into a transverse string of $2\ell+1$
lattice points, on which $|\Bphi|$ varies from nearly zero (in its center) to the surrounding vacuum 
configuration $\Bphi_0=(0,0,1)$, i.e.   $\Bphi=(0,0, |i|/\ell)$ for $-\ell\leq i\leq\ell$, on that string
of lattice points. Then we find for the contributions of a single soliton
\be\label{shrilat}
L^{(2)}_\bot\sim\frac{\tau}{\ell},~~~~~~~~~~ L^{(2)}_\|\sim\frac{\ell}{\tau},
 ~~~~~~~~~~U\sim\lambda\frac{\tau}{\ell}.
\ee
So, apparently, solitons shrinking on a lattice contribute to the energy terms which rise linearly with
proper time $\tau$ which (as lattice artifact) will dominate the total energy for large $\tau$.

We expect the winding density of the squeezed defect to be located on $\nu$ lattice squares 
near its center. This implies for the fourth-order term
\be\label{e4lim}
L^{(4)}_\|=\frac{\lambda\ell^2/\nu}{\tau}.
\ee
The winding density is determined by the orientation of the field unit-vectors alone, so it sufficient
to consider the unit-vectors $\hphi$. We expect
the squeezed defect to consist of just one unit-vector $\hphi=(0,0,-1)$ at the soliton center  
looking into the direction opposite to all surrounding unit-vectors $(0,0,1)$. That lattice point is the
top of four adjacent rectangular triangles (with the diagonals connecting the four nearest neighbour points as 
bases) which together cover an area of two lattice squares. So we expect
a winding density $\rho=1/\nu$ with $\nu\sim 2$. 
 
\begin{figure}[h]
\centering
\includegraphics[width=11cm,height=17cm,angle=-90]{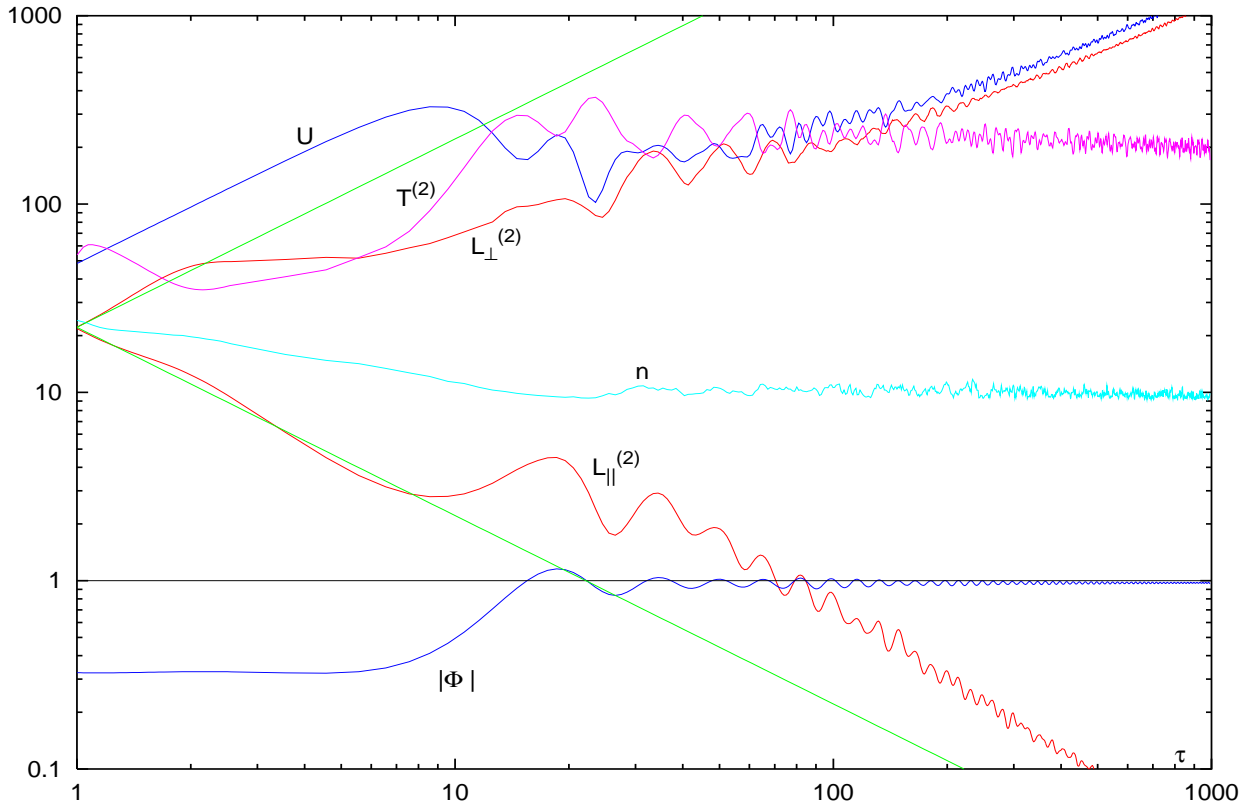}
\caption{\label{fig2} Potential energy $U$, kinetic energy $T^{(2)}$, transverse and longitudinal (second-order)
gradient terms $L_\bot^{(2)}$ and $L_{||}^{(2)}$, the number of defects $n$, and the average length of the 
chiral field $|\Bphi|$, for a typical evolution after a sudden quench (for $\lambda=1, 
\ell/a=4, \sigma_0=0.2, H=0.2, a\Te=0.2, N=50$). For comparison, the straight lines given by eqs.(\ref{linear}) 
with  (\ref{C2}) are included.} 
\end{figure}

This dominance of lattice artifacts for $\tau \gg \ell $ is illustrated in fig.(\ref{fig2})
which shows a typical evolution on a $50\times 50$ lattice for $\ell=4$.
The total winding number is $B=-1$. After the roll-down the number of solitons stabilizes at $n=9$
(cf. fig.\ref{fig1}).
Apparently, both $U$ and $L^{(2)}_\bot$,  approach a linearly rising limit for $\tau \gg 10^2$,
approximately like $\sim \frac{1}{2}n (\tau/\ell)$ which dominates the total energy, but does not
affect the (essentially constant) kinetic energy. Longitudinal contributions drop off like $\tau^{-1}$,
so they are irrelevant. 

It appears from fig.(\ref{fig2}) that for this evolution the roll-down (where the average of $\Phi$
aproaches the vacuum value $\Phi=1$) takes place during the time interval $2\ell <\tau < 4 \ell$,
i.e. long before artificial lattice effects dominate the energy. It is also by the end of the roll-down
that the number of created defects stabilizes. So we would conclude that results
obtained from lattice simulations for baryon-antibaryon production during the chiral phase transition 
in a rapidly expanding chiral gas are not severely affected by lattice artifacts.
On the other hand, to follow the evolutions beyond the end of the roll-down, which comprise
small '$\sigma$' and '$\pi$'-oscillations of $\Bphi$ around the true vacuum, interfering with small oscillations
of the bag profiles (resonances), will require to subtract the lattice artifacts.    

\section{Evolution until freeze-out}

In this chapter we will follow typical evolutions of the chiral field after a sudden quench 
in more detail and try to analyze their characteristic features up to the end of the roll-down.

Immediately before the sudden quench at $\tau=\tau_0$ the initial ensemble is prepared as described
in section II. The average length of one component of the chiral field is given by $\sigma_0$ (cf. eq. (\ref{fluc})),
the potential in (\ref{pot}) is characterized by a negative value of $f^2$. 
So, for sufficiently small $\sigma_0^2$ we have at $\tau=\tau_0$
\be \label{Uth0}
U_0= \tau_0\frac{\lambda}{4\ell^2} \int\left(f^4+2 |f^2|\langle\langle\Bphi^2\rangle\rangle\right) 
dx\:d\eta =(C_0+C_2){\cal V}_0,
\ee
where ${\cal V}_0=\tau_0\Delta\eta {\cal A}$ is the inital volume of the Bjorken slice, and the constants are
\be\label{const1}
C_0=\frac{\lambda}{4\ell^2}f^4, ~~~~~~~~~~~~C_2=\frac{3\lambda}{2\ell^2}|f^2| \sigma_0^2.
\ee
For the derivatives at lattice points $(i,j)$ Eq.(\ref{Fourier}) implies
\be  \label{deriv}
(\partial_x\Bphi)_{ij}=\frac{\mbox{i}}{N}\sum_{k,l=-N/2+1}^{N/2} \left(\frac{2\pi k}{aN}\right)
e\:^{\mbox{\small i}\frac{2\pi}{N}(i\cdot k +j\cdot l)} \tilde{\Bphi}_{kl},
~~~~~~(\partial_\eta\Bphi)_{ij}=\frac{\mbox{i}}{N}\sum_{k,l=-N/2+1}^{N/2} \left(\frac{2\pi l}{bN}
\right) e\:^{\mbox{\small i}\frac{2\pi}{N}(i\cdot k +j\cdot l)} \tilde{\Bphi}_{kl}.
\ee
Again replacing the integrands in (\ref{int3}),(\ref{int4}) by ensemble averages leads to
the second-order gradients contribution at $\tau=\tau_0$
\be \label{L20}
L_0^{(2)}=C^{(2)}\;{\cal V}_0.
\ee
For the constant $C^{(2)}$ we have in the continuum limit
\be \label{C2}
C^{(2)}=\frac{9}{2}\sigma_0^2 \Te^2 \left(1+\frac{m^2}{3\Te^2(1+\frac{m}{\Te})}\right).
\ee
Similarly, one may obtain a rough estimate for $L^{(4)}$ averaged over the initial ensemble
by replacing in (\ref{L4}) the unitvectors $\hphi$ by $\Bphi / \sigma_0$.

During the very early phase of an evolution in proper time 
the initially random ensemble of fluctuations will essentially stay random.
This means that the integrals in (\ref{int3})-(\ref{int5}) 
will remain constant, given by their initial values. 
Therefore, the time dependence of the different 
contributions (\ref{int3})-(\ref{int5}) to the total energy is given by the kinematical factors
($\tau/\tau_0$) or ($\tau_0/\tau$) alone, with the integrals approximated
by replacing the integrands through their averages in the initial ensemble.
     
After the quench, $f^2$ is positive, so for sufficiently small $\sigma_0^2$ we have
\be \label{Uth}
U= \tau\frac{\lambda f^2}{4\ell^2} \int\left(f^2-2\langle\langle\Bphi^2\rangle\rangle\right) 
dx\:d\eta =\frac{\tau}{\tau_0}(C_0-C_2){\cal V}_0,
\ee
and
\be \label{linear}
L_{\bot}^{(2)}=\frac{\tau}{\tau_0} \;C^{(2)}\;{\cal V}_0,~~~~~~~~~~~~~~~~~~~~~~~~~~~~~~~~
L_{\|}^{(2)}=\frac{\tau_0}{\tau}\;C^{(2)}\;{\cal V}_0.
\ee

In fig.(\ref{fig2}) both straight lines, (\ref{linear}) with (\ref{C2}), are included 
for comparison. It may be observed that the integral $L_{\|}^{(2)}$ involving the
longitudinal gradients follows the straight line decrease almost until the onset of the roll-down.
This means that the rapidity gradients basically stay random. On the other hand, the integral
$L_{\bot}^{(2)}$  follows the linear rise only for about one unit of proper time after the onset
of the evolution. Already near $\tau/\tau_0 \sim 2$, the transverse gradients 
are strongly affected by the dynamics  and interfere with the kinetic energy. 
Due to the relative factor of $1/\tau^2$ of $L_{\|}$ as compared to $L_{\bot}$ the dynamics
quickly gets dominated by the transverse gradients alone, such that
the average kinetic energy follows the average transverse-gradient energy $L^{(2)}_\bot$,
while the rapidity gradients (in $L_{\|}^{(2)}$ and $L_{\|}^{(4)}$) which decrease 
like $1/\tau$ are no longer relevant for the overall dynamical evolution.
 
Disregarding rapidity gradients altogether, the EOM (\ref{EOM}) reduces to
\be
\frac{1}{\tau}\partial_\tau\Bphi+\partial_{\tau\tau}\Bphi-\partial_{xx}\Bphi
 -m^2\Bphi=0,
\ee
which describes wave propagation in transverse direction, $A(\tau) \exp(ipx)$. 
Here the mass $m^2$ again characterizes the fluctuations around $\Bphi=0$, 
\be
m^2=\lambda f^2/\ell^2
\ee
so $m^2$ is negative for negative $f^2$ (where $\Bphi=0$ is the stable minimum),
and it is positive  for potentials which actually do have a 
lower symmetry-breaking minimum. The amplitudes $A(\tau)$
generically are Bessel functions,
\bea \label{waves}
A(\tau) &\sim& J_0(\tau\sqrt{p^2-m^2})~~~~~~~~~~~\mbox{for}~~~~ p^2-m^2 >0, 
\nonumber \\
A(\tau) &\sim& I_0(\tau\sqrt{m^2-p^2})~~~~~~~~~~~\mbox{for}~~~~ p^2-m^2 <0 .         
\eea 
For large values of their arguments the amplitudes of $J_0$ decrease like $1/\sqrt{\tau}$, 
while $I_0$ contains exponentially rising parts.
Modes with large transverse wave numbers contribute most to $L^{(2)}_\bot$.
Therefore, with their amplitudes decreasing like $1/\sqrt{\tau}$, the kinematical
factor $\tau$ in $L^{(2)}_\bot$ is compensated. So we expect that the linear rise of $L^{(2)}_\bot$ ends 
as soon as the dynamics is dominated by the transverse gradients and is followed by a phase where 
\be
\langle\langle \,T \,\rangle\rangle\sim\langle\langle L^{(2)}_\bot\rangle\rangle\sim \mbox{const.}|_\tau.
\ee
For negative $m^2$ no amplification occurs. After the quench, however, when $f^2$ has become positive,  
a few modes with small transverse wave numbers will start to get amplified.
Typically, for wave numbers $p=2\pi k/N$, with $k$ integer ($0\leq k\leq N/2$), 
waves with $k/N < \sqrt{\lambda}f/(2\pi\ell)$ get amplified, e.g. the lowest three or four out of $N=100$ for
$\ell\sim 5$ (for $\lambda=1$ and $f^2=1$). At first, the rate of amplification is slow because the exponential
rise is compensated by a decreasing function for small arguments in $I_0(x)$. These low-$k$ modes do not
contribute much to $L^{(2)}_\bot$. In fact, the $k=0$ mode, which experiences the largest rate of
amplification, does not contribute at all. 

While the amplification effect is not very pronounced for $L^{(2)}_\bot$, the few slowly exponentially 
rising contributions from the lowest-momentum transverse waves cause a noticeable rise of the
condensate $\langle\langle \Phi^2 \rangle\rangle$ after some time. This enters into the fluctuating part $C_2$ 
of the potential $U$ and drives it away from its linear rise given by (\ref{Uth}).
Then also the fourth-order terms in the potential become important and the dynamical evolution 
subsequently is dominated by the local potential. 
This initiates the roll-down of the field configuration at the majority of the lattice points into the
true vacuum $\Bphi_0=(0,0,1)$. The transition into the symmetry-violating configuration takes place,
with formation of bags and solitons in those regions where the winding density happens to be high.

To estimate the time $\tau_1$ for the onset of the roll-down we consider the $k=0$ mode with
amplitude $I_0(\tau m)$. Amplification of this amplitude by a factor $e$ in the time interval from
$\tau_0$ to $\tau_1$ requires
\be \label{tau1}
\ln I_0(\tau_1 m)= 1+ \ln I_0(\tau_0 m).
\ee
The r.h.s. depends only very weakly on $\tau_0$, as long as $(\tau_0 m)\leq 1$. In fact, $(\tau_1 m)$ varies
only from 2.26 to 2.55 for  $0\leq(\tau_0 m)\leq 1$. So, for convenience we simply take $(\tau_1 m)\approx 5/2$ if 
$(\tau_0 m)$ is of the order of 1 or less. Otherwise, for larger values of $(\tau_0 m)$, $\tau_1$ has to be obtained more
accurately from (\ref{tau1}). Typically, therefore, the transition from the gradient-dominated 
to the potential-dominated phase, happens near
\be\label{onset}
\tau_1 \sim \frac{5\,\ell}{2\sqrt{\lambda f^2}}.
\ee
However, up to this time $\tau_1$ of the onset of the roll-down, i.e. throughout the whole gradient-dominated phase 
the potential plays no significant role.
The overall evolution proceeds practically independently
from the (positive or negative) value of $f^2$ in the $\Phi^4$-potential (\ref{pot}). 
This also implies that the quenchtime
(the timescale for changes in $f^2$) is irrelevant as long as it is smaller than the time
during which the gradient terms dominate the evolution and it justifies the use of the sudden quench
approximation where we impose the 'cold' ($\Te=0$) potential from the outset at $\tau>\tau_0$.
 
With $\ell/\tau_0 > 1$, the ratio $(\tau_1/\tau_0)^2$ is sufficiently large to render all longitudinal
(rapidity) gradients unimportant as compared to the potential. 
This means that during the subsequent roll-down different rapidity slices become effectively decoupled,
and begin to evolve independently from each other, 
while in longitudinal directions the solitons contract to lattice unit size.
Within these rapidity slices, $\pi$- and $\sigma$-modes propagate transversely, 
and eventual further annihilations of soliton-antisoliton pairs take place
while the transverse shapes of the sqeezed bags are established.

By the end of the roll-down the remaining nontrivial and sufficiently separate structures
have essentially reached their stable form. Apart from small fluctuations, the integral 
(or sum) over the absolute values of the winding density $n=\int \!|\rho| \;dx dz$ then
stabilizes and counts the (integer) number of these defects. Therefore we identify
the end-of-the-roll-down time with the (chemical) freeze-out time $\tau_f$
when the numbers of baryons and antibaryons created are fixed. A rough estimate for $\tau_f$
may be obtained if we follow the further amplification of the amplitudes $I_0(\tau m)$ of the 
$k=0$ modes beyond $\tau_1$. For large arguments the increase in $I_0(\tau m)$ is mainly due to the
exponential $\exp(\tau m)$, so we obtain
\be\label{freeze}
\tau_f \sim \tau_1 + \frac{\ell}{\sqrt{\lambda f^2}}\ln\left(\frac{\Phi(\tau_f)}{\Phi(\tau_1)}\right)
\ee 
For a typical amplification ratio of 5 to 10 during roll-down we then find an approximate freeze-out time
of
\be\label{tauf}
\tau_f \sim \frac{4\ell}{\sqrt{\lambda f^2}}.
\ee
This certainly represents a lower limit for the duration of the
roll-down, because the increasing $\Phi^4$ contributions to the potential will slow down the
symmetry-breaking motion. The numerical simulations confirm this simple argument and indicate
that $(m \tau_f) \sim 4-5$ provides a reasonably accurate estimate for the freeze-out time
(as long as $(\tau_0 m) \leq 1$). 
After the quench, when $f_0^2$ has assumed its $(\Te=0)$-value $f_0^2=1$, we may neglect the 
small contributions of the explicit symmetry-breaking $H$ to $f^2$ and to the $\sigma$-mass
$m_\sigma^2$ in eqs.(\ref{f2}) and (\ref{masses}) and rewrite (\ref{tauf}) in the form
\be\label{taufreeze}
\frac{\tau_f}{\tau_0}\approx \frac{4\sqrt{2}}{\tau_0 m_\sigma}.
\ee

The typical example for an evolution given in fig.\ref{fig2} shows how during the roll-down
the configurations pick up an appreciable amount of kinetic energy  
until the potential starts to deviate from its linear rise and interferes 
with $\langle\langle \,T\rangle\rangle$. Subsequently, $\langle\langle U\rangle\rangle$ starts to pick up
the unphysical linearly rising lattice contributions (\ref{shrilat}) of the shrinking solitons, 
while the time-averaged $\langle\langle \,T\rangle\rangle$ remains basically constant. 
As the heavy solitons carry no kinetic energy, $\langle\langle \,T\rangle\rangle$ then resides
in small transversely propagating fluctuations which eventually are emitted as $\sigma$ and $\pi$ mesons.

\subsection{Correlation lengths and defect numbers}
In contrast to the integer net-baryon number $B=\int \!\rho \;d^2x$, the integral (or lattice sum) over the 
absolute values of the local winding density $|\rho|$ 
\be \label{defnum}
n=\int |\rho| \; dxdz
\ee
generally is not integer. 
The ensemble average of $n$ is closely related to the coherence  length $R$ for the 
field unit-vectors in the statistical ensemble of $O(3)$-field configurations. If an $O(n)$-field
is implemented on a $d$-dimensional cubic lattice with lattice constant $a$, then
the field orientations on the vertices of a sublattice with lattice unit $R/a$ can
be considered as statistically independent. Then the average 
$\langle\langle \; n \; \rangle\rangle$ expected on an $N^d$ lattice is 
\be\label{Kibble}
\langle\langle \; n \; \rangle\rangle=\nu_d \; (aN/R)^d
\ee
where $\nu_d$ is the average fraction of the surface of the sphere ${\bf S}^d$ covered by the image of the 
sublattice unit (this is the very definition of a winding density). 
The number $\nu_d$ can be estimated for different manifolds~\cite{Kibble}. 
For the map (compactified-)${\bf R}^2\rightarrow {\bf S}^2$
defined by the unit-vectors $\hphi(x,z)$ of the $O(3)$-field in $d=2$ dimensions it is $\nu_2=1/4$,
(i.e. $1/2^{d+1}$ for each of two triangles which make up each square sublattice unit cell). 

\begin{figure}[h]
\centering
\includegraphics[width=7cm,height=8.5cm,angle=-90]{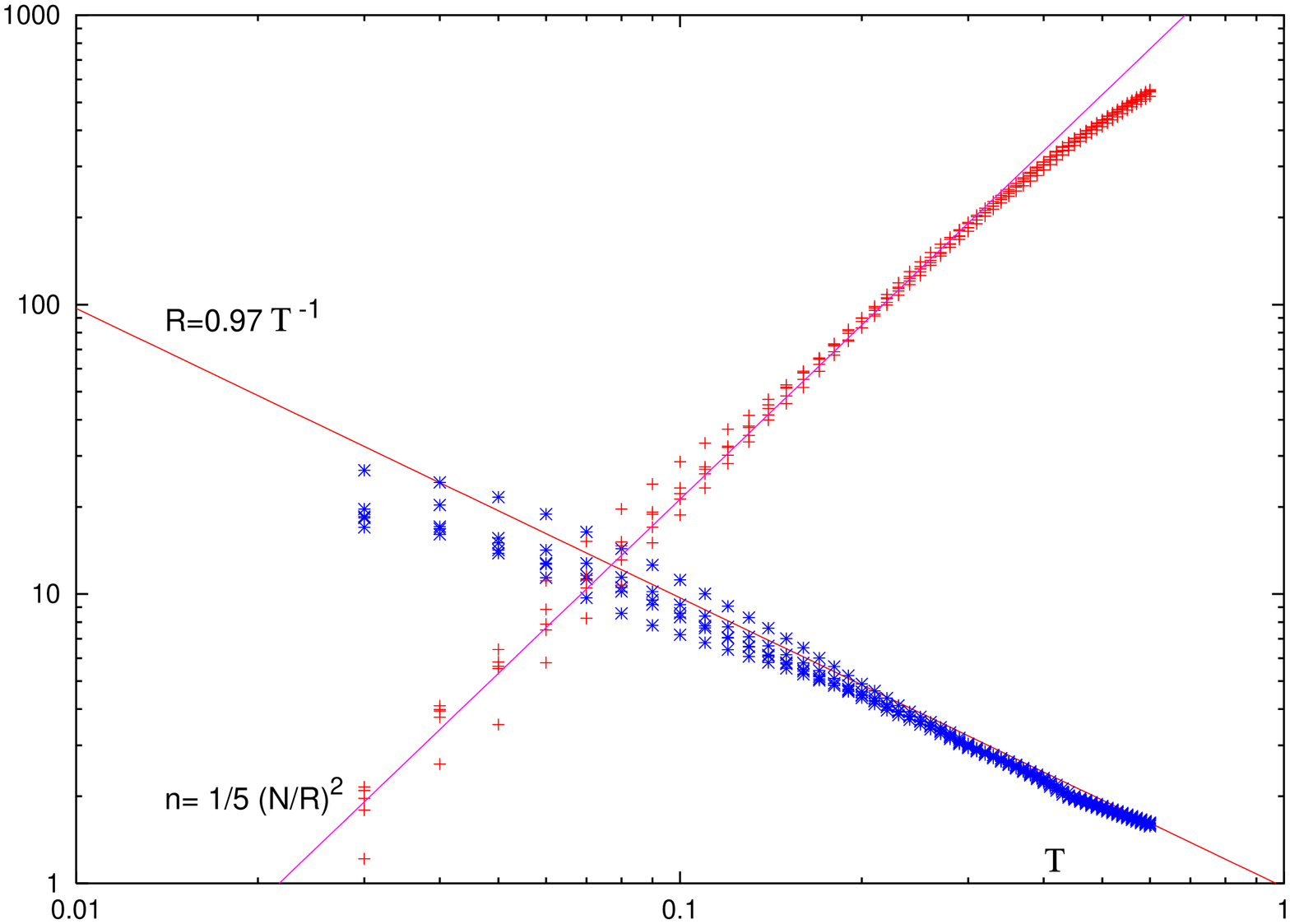}
 \includegraphics[width=7cm,height=8.5cm,angle=-90]{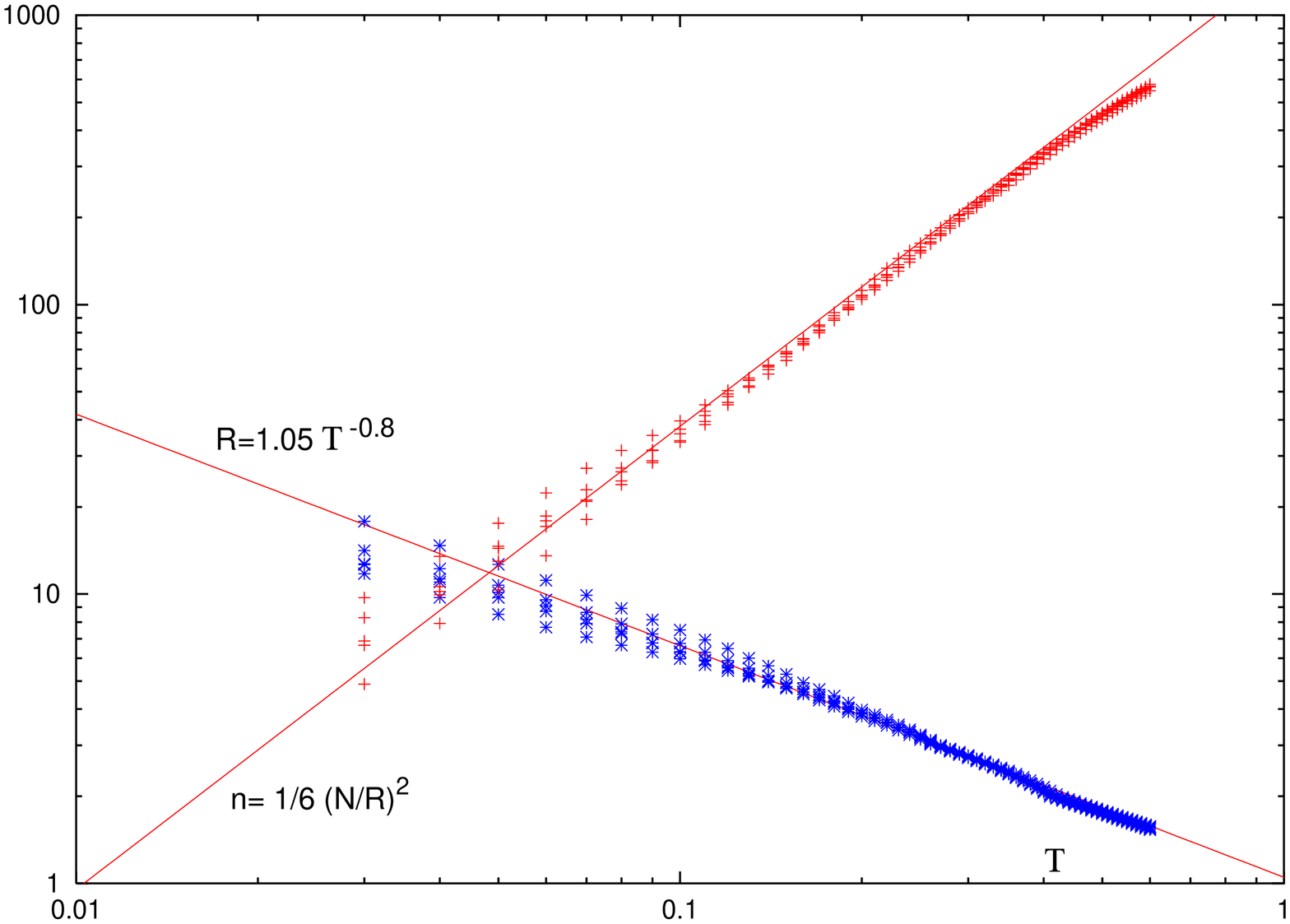}
\caption{\label{fig3}Initial (measured) coherence  lengths $R$ and number of defects $n$ 
as functions of initial temperature $\Te$, measured for five random initial
configurations for each temperature, on an $N=100$ lattice, with $m$ put to zero (left), 
and with $m=1/\ell$ for $\ell=4$ (right). (All quantities in lattice units $a$).}

\includegraphics[width=10cm,height=12cm,angle=-90]{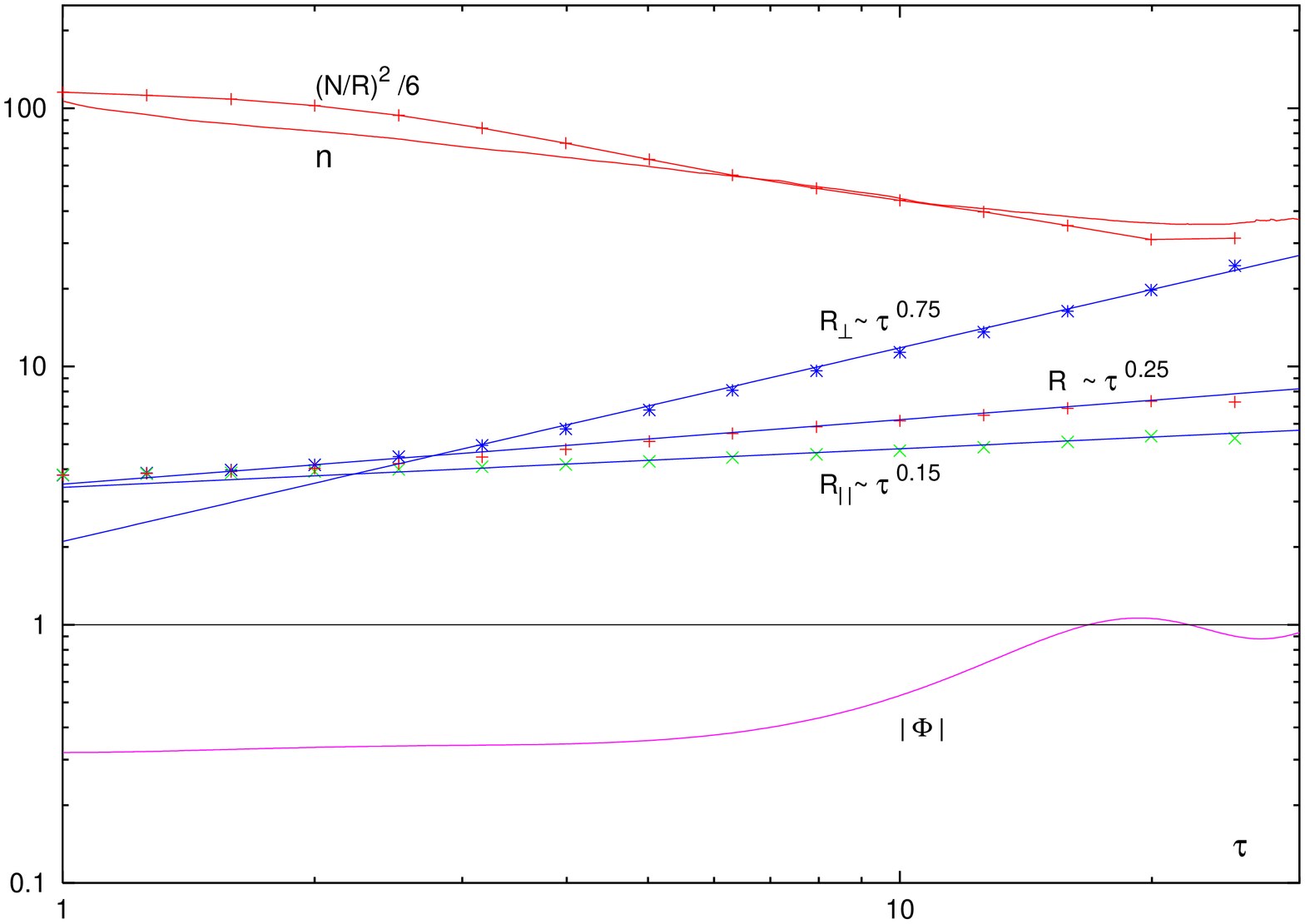}
\caption{\label{fig5} Crosses show the measured coherence  lengths $R_\bot, R_\|$ and the average ${\bar R}$ as obtained from
the definition (\ref{avcorr}), as functions of proper time $\tau$. For $\tau>2$ they are parametrized by
power laws with exponents 0.15, 0.75, and 0.25, respectively. The measured defect number $n$ (full line) 
is compared to the statistical result (\ref{Kibble}) with $\nu=1/6$ (crosses connected by lines).
($N=100$,$\Te=0.2$,$\sigma_0=0.2$,$\ell=4$,$H=0.1$,$\lambda=1$).}
\end{figure}

Inserting our result (\ref{Rm0}) (obtained for $m=0$) into this estimate for $d=2$ dimensions, leads to 
 $ \langle\langle \; n \; \rangle\rangle _{\tau=\tau_0} \propto \Te^2$.
 In fig.\ref{fig3}a this is compared with numbers $n$ and coherence  lengths $R$ measured 
 for several initial configurations on an $N\times N$
 lattice for different temperatures (for $N=100$ and mass $m=0$). Evidently, the finiteness of the lattice 
 causes a small systematic deviation from this $\Te^2$-dependence, especially for large values of 
 $\langle\langle \, n \, \rangle\rangle$, as the coherence  length approaches the lattice constant. 
 The measured numbers $\langle\langle \; n \; \rangle\rangle$ 
 follow (\ref{Kibble}) with satisfactory accuracy for $\nu\approx 1/5$. 
 In fig.\ref{fig3}b the same comparison is shown for non-vanishing mass $m=1/\ell$, for $\ell/a=4$.
 For $m\ne 0$ the coherence  length $R$ can be obtained from (\ref{corrinit}) and compared to the 
 measured values. Fig.\ref{fig3}b shows that they are reasonably well described
 by $R\propto\Te^{-0.8}$. The corresponding measured numbers $n$ follow (\ref{Kibble}) with good
 accuracy for $\nu\approx 1/6$. Of course, small changes of $\nu$ could be absorbed into a slightly
 redefined coherence  length (note that the correlation function (\ref{corrinit}) for $m=0$ does not decrease 
 exponentially). We shall, however, keep the definition (\ref{cut}).

The above considerations apply to random configurations which need not contain any
fully developed solitons but may consist of only small fluctuating local winding densities
which cover small fractions of the image sphere.   
However, if the configurations 
finally have evolved into an ensemble of well-separated solitons or antisolitons embedded in a 
topologically trivial vacuum with only small fluctuations in the local winding density, then the 
integral (\ref{defnum}) counts the number of these embedded baryons-plus-antibaryons. We therefore adopt the
notion 'number of defects' for $\langle\langle \; n \; \rangle\rangle$, irrespective whether
configurations comprise only small local winding densities, or partial or complete solitons.

For a typical evolution (see e.g. fig.\ref{fig2}) the number of defects
measured as function of proper time shows a slow decrease which follows approximately
a power law 
\be\label{powlaw}
n\sim n_{(\tau=\tau_0)}\left(\frac{\tau}{\tau_0}\right)^{-\gamma} .
\ee
By the end of the
roll-down at freeze-out time $\tau_f$ this decrease levels off and $n$ settles near the constant 
which counts the number of the finally surviving fully developed solitons-plus-antisolitons 
(cf. figs.\ref{fig1}). The decrease in $n$ reflects the slow increase
in the average coherence  length ${\bar R}$ up till the end of the roll-down. 
The longitudinal coherence  length $R_\|$ 
grows very slowly because rapidity gradients are suppressed with $1/\tau$ in the Bjorken frame. 
This leads to an effective decoupling of field-vectors in longitudinal direction
and subdues the drive for aligning field orientations in adjacent rapidity bins. On the other hand,
the transverse coherence  length $R_\bot$ grows rapidly. 
For $R_\bot\gg R_\|$, the average radius ${\bar R}$ obtained from (\ref{avcorr}) is dominated by $R_\|$.
A typical example is shown in fig.\ref{fig5} for an evolution which starts at $\tau_0/a=1$. 
The average ${\bar R}$ grows with an exponent of $\alpha \approx 0.25$. 
The statistical argument in (\ref{Kibble}) then leads to $n\sim \tau^{-2\alpha}$, with $2\alpha=\gamma\approx 0.5$. 
This is slightly steeper
than the measured decrease in $n$. But as the growth in the coherence  radii sets in only after one
or two units of $\tau$ after the onset of the evolution, the final number of surviving defects is
reasonably well reproduced by the statistical expression (\ref{Kibble}) 
(we adopt $\nu=1/6$ from fig.\ref{fig3}b). 
Altogether, we typically find exponents $\gamma\sim 0.4\pm 0.05$
for the decrease (\ref{powlaw}) of the number of defects. Then, with
 (\ref{taufreeze}) for the typical freeze-out time, we have
\be \label{decrease}
\langle\langle \; n \; \rangle\rangle_{|\tau=\tau_f}=\langle\langle \; n \; \rangle\rangle_{|\tau=\tau_0}
\left(\frac{\tau_0\;m_\sigma}{4\sqrt{2}}\right)^{0.4}
\ee
for the reduction of  the number of defects  from its initial value at the onset of the evolution until
the end of the roll-down. With $(\tau_0\,m_\sigma)$ of the order of 0.5 to 1, 
we find reduction factors
of 1/3 to 1/2, which is not even one order of magnitude. So, this is not a dramatic result. The reason
is, evidently, that in the expanding Bjorken frame 
the gradient coupling in rapidity direction quickly gets suppressed.

\begin{figure}[h]
\centering
\includegraphics[width=9cm,height=14cm,angle=-90]{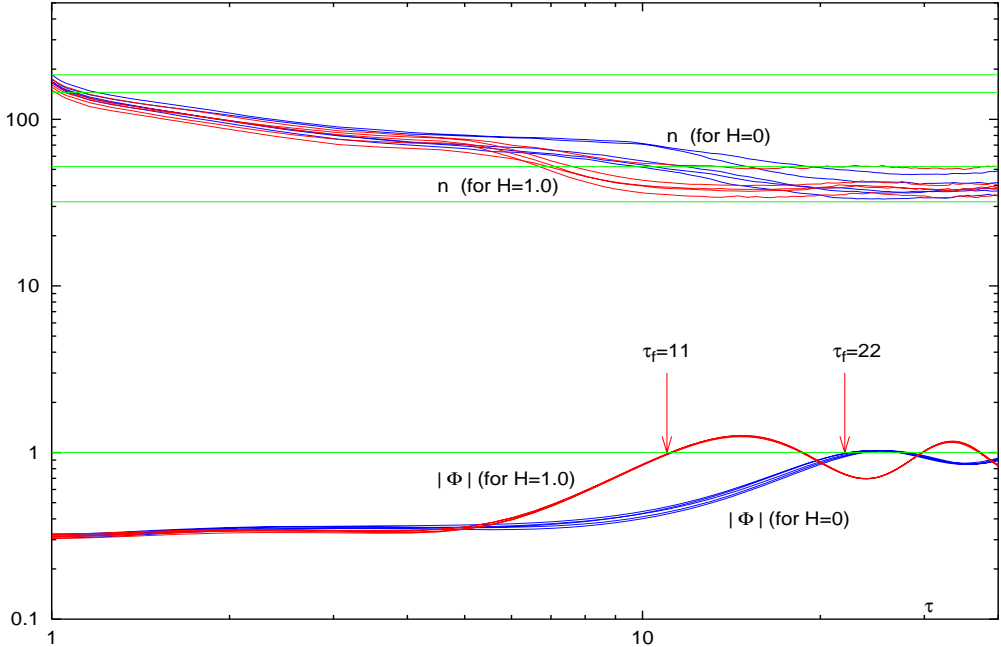}
\caption{\label{fig4}The numbers of defects $n$ for two different values $H=0$ and $H=1.0$ for the strength of the 
explicit symmetry breaking, each for five evolutions on a $100\times 100$ lattice ($\Te=0.3,\lambda=1,\ell=5,
\sigma_0=0.2$). The arrows point to the freeze-out times $\tau_f\approx 22$, and $\tau_f\approx 11$, respectively,
where the average lengths $|\Bphi|$ of the chiral field vectors have reached  $|\Bphi|=1$. 
The initial values of $n$ lie within a band from 145 to 185, 
they all end (for both values of $H$) in a band from 32 to 52, which corresponds to reduction
factors of about 1/4.  }
\end{figure}

It should be noted that all numerically measured exponents are independent of the
choice of the lattice constants, because scaling $x\rightarrow ax$, (i.e.$\ell\rightarrow a\ell$), 
$\eta\rightarrow b\eta$, $\tau\rightarrow (a/b)\tau$ leaves the EOM (\ref{EOM}) invariant.
The length unit $a$ only
serves to define the resolution with which the spatial structure of the field configurations is analyzed
and all physical results should be independent of this scale. On the other hand, the initial time $\tau_0$
denotes the physical point in time when the system begins its evolution in terms of hadronic degrees 
of freedom with a sudden or rapid quench in the relevant potential. So, physical  results generally will
depend on $\tau_0$, as is evident from the reduction factor obtained in (\ref{decrease}).

Small explicit symmetry breaking ($H\ne 0$) accelerates the decrease of $n$
during roll-down, but at the same time it reduces the freeze-out time, such that the final number of $n$
remains essentially unaffected by small non-zero values of $H$.
Fig.\ref{fig4} shows a number of evolutions for two different
strengths $H$ of explicit symmetry breaking. 

The same is true if additional damping is introduced into the 
EOM (\ref{EOM}) by adding a term $\kappa\partial_\tau\Bphi$ with damping constant $\kappa$ to account for the
fact that the field fluctuations are actually emitted from the expanding Bjorken rod, carrying away energy.
Through this dissipative dynamics the evolutions are slowed down, the roll-down times may be retarded by 
an order of magnitude, but the overall reduction factor in the number of surviving defects remains 
unaffected. Of course, all fluctuations then are damped away during the course of the evolution, and the integrals
(\ref{int3}) to (\ref{int5}) finally are determined by the remaining ensemble of squeezed solitons alone, 
while the kinetic energy goes to zero.

\subsection{Meson spectrum}

For times long after the roll-down the average kinetic energy  $\langle\langle \; T \; \rangle\rangle$,
and the potential energy parts $\langle\langle \; L_\bot+U \; \rangle\rangle$ (after subtraction of the
linearly rising (lattice) contributions from the squeezed solitons (\ref{shrilat})) converge towards the same constant
$E_f/2$. Their sum $E_f$ represents the average total energy stored in the mesonic field fluctuations 
after the roll-down. Both averages show residual fluctuations around their smooth background 
with opposite phases, such that their sum $E_f$ is smooth. Analysing the spectral density of  
either $\langle\langle \; T \; \rangle\rangle$, or $\langle\langle \; L_\bot+U \; \rangle\rangle$,
(after subtracting the background), tells us about the spectral distribution of pions and 
$\sigma$-mesons that will eventually be emitted from the expanding Bjorken rod. 

\begin{figure}[h]
\centering
\includegraphics[width=8cm,height=14cm,angle=-90]{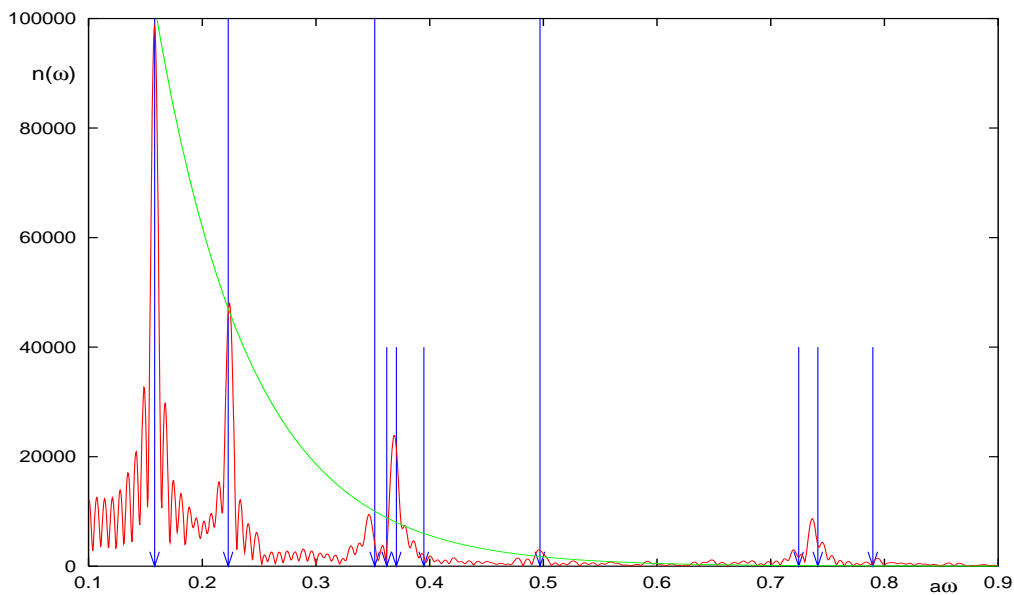}
\caption{\label{fig6} Spectral density $n(\omega)$ of the residual fluctuations in the 
average kinetic energy $\langle\langle \; T \; \rangle\rangle$ for times long after the roll-down,
($N=80$, $\ell/a=4$, $H=0.1$). The arrows point to the lowest (double-)frequencies (\ref{omega}) for $\pi$- and
$\sigma$-mesons (see text). The green line is the exponential $\exp(-12a\omega)$.}
\end{figure}

We consider the Fourier-transforms 
\be
c(\omega)+i s(\omega)=\int_{\tau_a}^{\tau_b} \langle\langle \; T(\tau) -{\bar T}(\tau)\; \rangle\rangle
 e^{i \omega \tau} d\tau,
\ee
where the integral covers times long after the roll-down, e.g. $\tau_a/\tau_0\sim 100 $, $\tau_b/\tau_0\sim 1000$,
and ${\bar T}(\tau)$ subtracts the smooth background. The absolute value, $\epsilon(\omega)= \sqrt{c^2+s^2}$,
represents a spectral energy density, from which we may extract the spectral particle number density 
\be
n(\omega)=\frac{\epsilon(\omega)}{\omega} = \sum_{ij}n_{ij}^{(\pi)}\;\delta(\omega-2\omega_{ij}^{(\pi)})+
\sum_{ij}n_{ij}^{(\sigma)}\;\delta(\omega-2\omega_{ij}^{(\sigma)}) + \cdot\cdot\cdot  .
\ee
The $ij$-sum with $i,j=0,1,2,...N/2$ covers all frequencies on the lattice for pions and 
$\sigma$-mesons with masses $m_\pi$ and $m_\sigma$ given in (\ref{masses})
\be \label{omega}
\omega_{ij}^{(\pi/\sigma)}=\sqrt{\left(\frac{2\pi}{aN}i\right)^2+
\left(\frac{2\pi}{a N}j\right)^2 + m_{\pi/\sigma}^2} ~~.
\ee  
Generically, $T(\tau)$ contains contributions $\sim (\cos(\omega_{ij}^{(\pi)}\tau))^2$ from the pionic
fluctuations, and $\sim (\cos(\omega_{ij}^{(\sigma)}\tau)+ c)^2$ from the $\sigma$-fluctuations around some
nonvanishing average $c$. Therefore, the spectral functions $\epsilon(\omega)$ and $n(\omega)$ will, in addition
to the double frequencies $2\omega_{ij}^{(\sigma)}$, also contain contributions for the $\sigma$-mesons 
at the single frequencies $\omega_{ij}^{(\sigma)}$.

Figure \ref{fig6} shows the spectral density $n(\omega)$
as obtained from the residual fluctuations in the average kinetic energy. 
The long vertical arrows point to the first four $2\omega_{ij}^{(\pi)}$
pionic frequencies (\ref{omega}) for $ij=00,10,20,30$, with $m_\pi^2=H/\ell^2$, ($H=0.1,\ell/a=4,f_0=1$).
It may be seen that the overwhelming part of the strength resides in the lowest and first excited pionic
modes. The strength decreases rapidly with excitation energy, approximately like $\exp(-12a\omega)$. 
The same is true for the strength of the $\sigma$-modes. (The short arrows in fig.\ref{fig6}
point to the first three modes
with $ij=00,10,20$, with single frequencies $\omega_{ij}^{(\sigma)}$ and double frequencies 
$2\omega_{ij}^{(\sigma)}$). However, the number density $\sum n_{ij}^{(\sigma)}$ for the sigmas, 
which we may extract from the strength located at the double frequencies
$2\omega_{ij}^{(\sigma)}$, is only about $5\%$ of the pionic strength residing in the first three pionic modes.
For an order-of-magnitude estimate of the pionic multiplicities we therefore ignore the $\sigma$-contributions.

\subsection{Meson and baryon multiplicities}

To obtain a simple estimate for the energy $E_f$ finally available for meson production
we consider the time of the onset of the roll-down $\tau_1$ in (\ref{onset}) which marks the transition from 
the gradient-dominated to the potential-dominated phase. At this time, for sufficiently small $\sigma_0^2$, 
the total energy is dominated by the linearly rising term $(\tau/\tau_0)C_0 {\cal V}_0$ in the potential
(\ref{Uth}). With the onset of the roll-down the average potential 
$\langle\langle \; U \; \rangle\rangle$ starts to deviate from this linear rise and bends
down to interfere with $\langle\langle \; T \; \rangle\rangle$ and $\langle\langle \; L \; \rangle\rangle$,
(cf. fig. \ref{fig2}). In the numerical simulations the large-time limit of 
$\langle\langle \; U \; \rangle\rangle$ and $\langle\langle \; L \; \rangle\rangle$ is masked 
by the (lattice-artificial) rise of the soliton contributions. But the asymptotic 
$\langle\langle \; T \; \rangle\rangle$ is free of these artifacts and (apart from residual fluctuations)
approaches a constant value, which  
is well represented by the linearly rising $\frac{1}{2}(\tau/\tau_0)C_0 {\cal V}_0$ taken at $\tau=\tau_1$. 
Approximating $\tau_1$ by $5/(\sqrt{2}m_\sigma)$ as given in (\ref{onset}), we then have (with $f^2=1$)
\be \label{Ef}
E_f = f_\pi^2\frac{\tau_1}{\tau_0}C_0{\cal V}_0\approx f_\pi^2\frac{5 m_\sigma}{8\sqrt{2}}(ab N^2),
\ee
where we have again neglected the small contribution of the explicit symmetry-breaking $H$ to the
$\sigma$-mass.
Within this level of accuracy we can also ignore that about $30\%$ of the pions 
carry the energy $\omega_{10}^{(\pi)}$ (instead of $m_\pi$), 
and obtain the pion multiplicity $n_\pi$ from dividing (\ref{Ef}) by $m_\pi$,
\be
n_\pi=\frac{5}{8\sqrt{2}}\frac{m_\sigma}{m_\pi} f_\pi^2 (ab N^2).
\ee
This number may be compared with the baryon-plus-antibaryon multiplicity given in (\ref{decrease}). 
We use for $\langle\langle \; n \; \rangle\rangle_{|\tau=\tau_0}$ the statistical result (\ref{Kibble}),
with initial spatial coherence  length $R_0$. Then we have
\be
\langle\langle \; n \; \rangle\rangle = \nu\left(\frac{aN}{R_0}\right)^2
\left(\frac{\tau_0\;m_\sigma}{4\sqrt{2}}\right)^\gamma .
\ee 
The last factor relies on the estimates (\ref{onset}) and (\ref{tauf}) for the times $\tau_1$ and $\tau_f$,
which are valid as long as $(\tau_0 m)\leq 1$, (otherwise they have to be obtained more accurately from
(\ref{tau1}) and (\ref{freeze})).
The evolutions described above have been performed for initial configurations selected with netbaryon
number $B=0$. So, the average number $n_{\bar p}$ of antibaryons created during the phase transition 
is $\langle\langle \; n \; \rangle\rangle/2$. With typical values $\nu\sim 1/4$, $\gamma\sim 0.4$
we find for the multiplicity ratio of antibaryons to pions
\be
n_{\bar p}/n_\pi \approx 0.14 \frac{a}{b}\frac{m_\pi }{m_\sigma f_\pi^2}\frac{(\tau_0 m_\sigma)^\gamma}{R_0^2}.
\ee
With an overall energy scale $f_\pi^2$ of the order of the pion mass $m_\pi$, and $R_0$ of the
order of $m^{-1}=\sqrt{2} m_\sigma^{-1}$, this ratio is 
\be
n_{\bar p}/n_\pi \approx 0.07 (\frac{a}{b}m_\sigma) (\tau_0 m_\sigma)^\gamma .
\ee
The ratio $a/b$ of the spatial and rapidity lattice constants which appears in this result has a 
physical meaning: according to (\ref{initcorr}) it is equal to the ratio of the (transverse) spatial 
coherence  length $R_0$ and the (longitudinal) rapidity coherence  distance $R_{\| 0}$
in the initial configuration. Naturally, this ratio is of the order of $\tau_0$.
 So, for initial times $\tau_0$ typically of the order of the inverse
$\sigma$-mass we find antibaryon-to-pion multiplicity ratios of the order of $0.05$ to $0.1$.

\section{Generalization to 3-d O(4)}

For the generalization to the 3+1 dimensional O(4)-model we keep the parametrization as given in eqs.(\ref{lag}),
(\ref{L2}), and (\ref{pot}). In this case $f_\pi^2$ is an overall constant of dimension [mass$^2$], 
so the physical fields $f_\pi\Bphi$ are of mass-dimension one.
The winding density is no longer given by (\ref{top}),
but we keep ${\cal{L}}^{(4)}$ as defined by the second equality in eq.(\ref{L4}). Conventionally, the strength of
the  ${\cal{L}}^{(4)}$-term in (\ref{L4}) is given in terms of the Skyrme parameter $e$ as
\be\label{Skyrme}
\frac{2\lambda\ell^2}{(8\pi)^2} \Rightarrow \frac{1}{4e^2f_\pi^2}.
\ee
In this case the typical spatial radius of a stable skyrmion in its rest frame is mainly determined
by the balance between ${\cal{L}}^{(2)}$ and ${\cal{L}}^{(4)}$, so it is of the order of $(ef_\pi)^{-1}$.

For the map (compactified-)${\bf R}^3\rightarrow {\bf S}^3$ defined by the unit vectors of 
the $O(4)$-field in 3 spatial dimensions 
the statistical result (\ref{Kibble}) for the average number of defects found on a $(aN)^3$ lattice 
for initial configurations with coherence  length $R_0$ generalizes as
\be\label{Kibble3}
\langle\langle \; n \; \rangle\rangle|_{\tau=\tau_0}= \frac{5}{2^{3+1}}\; (aN/R_0)^3.
\ee
(We again use $a=b\tau_0$ for the lattice constants).
The factor 5 counts the number of 3-simplices (tetraeders) which make up a cubic sub-lattice cell of size $R_0^3$,
the factor $(1/2^{d+1})$ with $d=3$ is the (absolute value of the) average surface area  covered by the image
of one 3-simplex on the image sphere ${\bf S}^3$.  So the factor $5/16$ counts the average 'number of defects'
associated with a cubic lattice cell with lattice constant given by the initial coherence  length $R_0$.  
A certain arbitrariness in the
definition of the coherence  length may translate into modifications of this factor $5/16$;
(e.g. for a random lattice of 3-simplices the factor 5 is replaced by ($24\pi^2/35 \sim 6.8$) \cite{Kibble}).
In any case we do not expect order-of-magnitude changes in this factor as compared to the $d=2$ case, where we
had $2/2^{2+1}$.

However, through the cubic power the result is now very sensitive to the actual value of $R_0$ in the
initial ensemble. Different concepts about the physical nature of the initial configurations will imply 
quite different ways to arrive at the appropriate initial coherence  lengths $R_0$. 
For an initial ensemble which is characterized by a temperature $\Te$ we could proceed
as in (\ref{corrinit}) and relate $R_0$ to the temperature, or to the mass 
$m^2(\Te)=\lambda |f^2(\Te)|/\ell^2$ of the field fluctuations;
but it has also been suggested \cite{Ellis}
to tie $R_0$ to the parton density which makes it independent of the temperature concept.
So, for the moment it seems appropriate to keep the initial coherence  length $R_0$ as a parameter.   

Adding a second transverse dimension does not change the result (\ref{avcorr}) for the average of the
transverse and longitudinal coherence  lengths. The growth in the resulting ${\bar R}$ again is dominated 
by the slow increase of the longitudial coherence  length $R_\|$ which is unaffected by additional
transverse dimensions. The estimates (\ref{onset}) and (\ref{tauf}) for the times $\tau_1$ 
and $\tau_f$ of the onset and end of the roll-down also remain unaffected, as they only rely on 
the amplitudes $A(\tau)$ of the transverse waves (\ref{waves}), 
irrespective of the number of spatial dimensions. (In this case, ${\cal{L}}^{(4)}$ now
contributes to $L_\bot$ with a term containing four transverse gradients, acting on the direction of the
O(4) field. The roll-down, however, takes place in areas which are topologically trivial, i.e. with small
angular gradients, so we do not expect a strong effect on the roll-down times. 
Within the approximations which led to (\ref{decrease}), we then find
for the average number of baryons+antibaryons present after the roll-down
\be\label{defects}
\langle\langle \; n \; \rangle\rangle_{|\tau=\tau_f}=\frac{5}{16}\; \left(\frac{a N}{R_0}\right)^3
\left(\frac{\tau_0\;m_\sigma}{4\sqrt{2}}\right)^{3\alpha}
\ee
with $\alpha\sim 0.2$ to 0.25. 
We denote the transverse area $(aN)^2$ of the Bjorken rod by ${\cal A}$, and replace the ratio $(a/b)$ 
of the lattice constants again by the initial time $\tau_0$. Then we obtain for the 
rapidity density of antiprotons ($n_{\bar p}=\frac{1}{4}\langle\langle \; n \; \rangle\rangle$)
\be\label{pbar}
\frac{d n_{\bar p}}{d \eta}=\frac{5}{64}\; \frac{\tau_0{\cal A}}{R_0^3}
\left(\frac{\tau_0 m_\sigma}{4\sqrt{2}}\right)^{3\alpha}.
\ee
(At this point we count all baryons as nucleons, assuming that excitational fluctuations and rotations
contribute to the surrounding pionic fluctuations).
Although the strength of the Skyrme term does not appear explicitly in (\ref{pbar}), the presence of the
${\cal{L}}^{(4)}$-term is essential for the formation of the solitons, because during the evolution 
it transforms the
sum over the absolute values of average winding densities into the average number of fully developed
solitons and antisolitons. So it is hidden in the growth law characterized by $\alpha$. 

For meson production we adopt the considerations which led to the estimate (\ref{Ef}). Counting again all
mesons as zero momentum pions we have 
\be \label{npi}
n_\pi=\frac{E_f}{m_\pi} = \frac{f_\pi^2}{m_\pi}\frac{\tau_1}{\tau_0}C_0{\cal V}_0
=f_\pi^2\frac{5}{8\sqrt{2}}\frac{m_\sigma}{m_\pi}({\cal A}\,b N).
\ee
The rapidity density of negatively charged pions $n_{\pi^-}=\frac{1}{3}n_\pi$ then is
\be \label{piminus}
\frac{d n_{\pi^-}}{d \eta}= f_\pi^2\frac{5}{24\sqrt{2}}\frac{m_\sigma}{m_\pi}{\cal A}.
\ee
In a heavy-ion collision the transverse area ${\cal A}$ of the Bjorken rod will correspond to the 
spatial overlap of the colliding relativistic nuclear slabs. As we have assumed spatially homogeneous 
initial conditions we have to consider slabs with constant nucleon (area-)density. In order to account for the 
number A of nucleons contained in one slab, its radius must be taken as $r_0$A$^{1/2}$,
with $r_0\approx 1.2$ fm. Then, as function of centrality, $(d n_{\pi^-}/d \eta)$ is directly 
proportional to the number of participants $N_p$, which is one of the basic experimental results in
relativistic heavy-ion collisions. For central collisions of A-nucleon slabs we have $N_p = 2$A, so we find
for the $\pi^-$-rapidity density per $N_p/2$ participants 
\be \label{ppp}
\frac{1}{N_p/2}\frac{d n_{\pi^-}}{d \eta}= \frac{5 \pi}{24\sqrt{2}}\frac{m_\sigma}{m_\pi}(r_0 f_\pi)^2.
\ee
This is an interesting result because all parameters have been absorbed into physical quantities. 
There are, however, several caveats: We have used for this result the form of the potential (\ref{pot}) {\it after} 
the quench (only this enters into the calculations). This means that differences between the average potential energy 
{\it before} the quench (\ref{Uth0}) and immediately {\it after} the quench (\ref{Uth}) are left out. 
However, this difference is of the order $\sigma_0^2$, 
which has been neglected in (\ref{ppp}) anyway. But generally, the result (\ref{ppp}) should be 
considered as a lower limit. It should further be noted
that the result (\ref{ppp}) depends linearly on the time $\tau_1$ for the onset of the roll-down.
The definition of $\tau_1$ in (\ref{onset}) is not very stringent and may be subject to changes by
$\pm 20\%$. The estimate (cf. eq.(\ref{onset})) we used for the time $\tau_1$ 
required $(\tau_0 m_\sigma)\leq\sqrt{2}$ , 
i.e. with $m_\sigma \sim 3-5$ fm$^{-1}$, the initial time $\tau_0$ should not exceed $0.3-0.5$ fm.
Another unsatisfactory feature of the homogeneous Bjorken rod is that the inhomogeneity
in the nucleon (area-)density of real relativistic nuclear slabs with transverse radii $r_0$A$^{1/3}$ 
has to be represented through radii $r_0$A$^{1/2}$ for homogeneous slabs.

The experimental value \cite{phenix} for the $\pi^-$-rapidity density per $N_p/2$
lies between 1.2 and 1.5 for $N_p$ increasing up to 350.
With $r_0 f_\pi=0.57$, $m_\sigma/m_\pi\approx 5-8$, our result (\ref{ppp}) leads to
\be \label{pipp}
\frac{1}{N_p/2}\frac{d n_{\pi^-}}{d \eta}\approx 0.75-1.2 ~.
\ee
 In the light of the reservations discussed above, this is quite satisfactory.
 
 In contrast to the parameter-free pion multiplicities, the result for baryon-antibaryon creation
depends on two parameters: the time $\tau_0$ when the initial hadronic field ensemble is established 
and begins its expansion, and the initial coherence  length $R_0$ within that ensemble.
 From  (\ref{pbar}) and  (\ref{piminus}) we have (with $\alpha\sim 0.2$ to 0.25)
\be\label{ratio}
\frac{n_{\bar p}}{n_{\pi^-}}\approx 
0.15 \left(\frac{m_\pi m_\sigma}{f_\pi^2}\right)\frac{(\tau_0 m_\sigma)^{3\alpha+1}}{(R_0 m_\sigma)^3}.
\ee
The experimental value for the ratio of integrated ${\bar p}$ to $\pi^-$ multiplicities 
lies between 0.065 and 0.085 \cite{phenix} for varying numbers of participants.
With $m_\pi/f_\pi^2=3.0$ fm, and a typical $\sigma$-mass of $m_\sigma \approx$ 3 fm$^{-1}$,
the experimentally observed multiplicity ratios are reproduced if $R_0$ and $\tau_0$ (both in [fm]) satisfy
\be \label{rule}
R_0 \approx (3 \tau_0)^{\alpha+1/3}.
\ee
For initial times in the range  0.2 $\leq \tau_0\leq$ 0.5 the dependence on $\alpha$  
is very weak and the coherence  length varies in the range 0.7  $\leq R_0 \leq$ 1.2 (all in [fm]).
These values are certainly within the limits of conventional assumptions. Interpreted in terms of
a thermodynamic equilibrium ensemble, $R_0\sim 1\mbox{fm}\sim \Te^{-1}$ implies the standard
estimate $\Te\sim 200$ MeV for the chiral phase transition. With our choice $a/b=\tau_0=R_0/R_{\| 0}$ for the ratio
of the spatial and rapidity lattice constants, the initial time $\tau_0 \approx 1/3$ fm resulting from (\ref{rule})
for $R_0=1$ fm then means that in the initial ensemble the initial rapidity coherence  distance $R_{\| 0}$ 
extends over three units of rapidity.
      
\section{Conclusion}

We have presented numerical simulations of the dynamical evolution which chiral field configurations
undergo in a rapidly expanding spatial volume. Starting at an initial time $\tau_0$ from a random hadronic
field ensemble with restored chiral symmetry, we follow its ordering process and roll-down into the
global potential minimum with spontaneously broken chiral symmetry. In accordance with standard concepts of
heavy-ion physics we have considered one-dimensional longitudinal expansion of an essentially baryon-free
region of high energy density, as it may be realized in the aftermath of an ultra-relativistic collision 
of heavy ions for central rapidities. 

Performed on a space-rapidity lattice in proper time of comoving frames, 
such simulations are very powerful instruments which
allow to investigate a multitude of interesting features related to the chiral phase transition.    
We have concentrated here on the topological aspects which are directly related to baryon-antibaryon
multiplicity as a sensitive signal for the phase transition. Mesonic abundancies could be analysed as well,
both for $\pi$ and $\sigma$ mesons (or any other elementary fluctuations included in the chiral field).
Not only their spectra can be obtained, but from the instantaneous configurations the spectral power of their
momentum distribution could be extracted at every point in time.

The method is not restricted to thermally equilibrated initial ensembles  with global or local temperature;
inhomogenities and anisotropy in the correlation lengths could be implemented naturally. Surface effects could be 
investigated by suitable boundary conditions. This may be interesting with respect to the $A$-dependence of
spectra and multiplicities. Here we have applied only standard periodic conditions. 
The one-dimensional expansion could be replaced by anisotropic or spherically symmetric expansion, which may be
of specific interest in cosmological applications.  
We have used the sudden quench approximation, which could be replaced by any desired time-dependence of
the chiral potential with arbitrary quench times. 
We have selected ensembles with conserved net-baryon number $B=0$ or very small $B$. 
Any other choice would be possible, and it appears as a peculiarly attractive feature to
study evolutions in ensembles with high net-baryon density, either fixed or in the form of grandcanonical
ensembles. The method is well suited to analyse distribution, growth and realigning of domains with 
disoriented chiral condensate as has been shown previously in purely dissipative dynamics~\cite{HoKl02}.
The generalization to SU(3)-fields appears most interesting, to learn about strangeness production
in terms of baryonic and kaonic abundance ratios.  
 
Evidently, the method opens up a wide field of applications. Unfortunately, however, we know very little
about the nature and characteristics of the initial ensemble which enters crucially into all physical results.
So, in our present analysis of antibaryon and pion multiplicities, the experimental data do not
allow to draw definite conclusions about the validity of the topological approach, because the results depend 
on two initial coherence lengths, the spatial $R_0$ and rapidity $R_{\| 0}$, 
(which for an isotropic initial ensemble are related by $R_0=R_{\| 0}\tau_0$). We can only conclude that
conventional assumptions about these quantities lead to results which are compatible with experimentally 
detected multiplicities. So, luckily, the mechanism is not ruled out. On the other hand,
an assumption like $\tau_0=R_0$ ( which would imply that the correlations have
grown with the speed of light from a pointlike origin ) is ruled out: it would overestimate the abundance
ratio in (\ref{ratio}) by a factor of 5.

\section{Acknowledgement}
The author appreciates helpful discussions with J. Klomfass, H. Walliser, and H. Weigel.


\begin{thebibliography}{99}
\itemsep=-0.1cm

\bibitem{Skyrme} T.H.R. Skyrme, {\it{Proc.R.Soc.}} {\bf A260} (1961) 127;\\
 
 \bibitem{Schwesinger} B. Schwesinger, {\it{Nucl.Phys.}} {\bf A537} (1992) 253;\\
 H. Weigel, {\it{Int.J.Mod.Phys.}} {\bf A11} (1996) 2419;\\
 Y.S. Oh, D.P. Min, M. Rho, and N.N. Scoccola, {\it{Nucl.Phys.}} {\bf A534} (1991) 493;\\

\bibitem{Nakano} M.Chemtob, {\it{Nucl.Phys.}} {\bf B256} (1985) 600;\\
H. Walliser, {\it{Nucl.Phys.}} {\bf A548} (1992) 649;\\
D. Diakonov, V. Petrov, and M.V. Polyakov, {\it{Z.Phys.}} {\bf A359} (1997) 305;\\
H. Weigel, {\it{Eur.Phys.J.}} {\bf A2} (1998) 391;\\
T. Nakano et al., [LEPS Collaboration], {\it{Phys.Rev.Lett.}} {\bf 91} (2003) 012002;\\


\bibitem{Hayashi} A. Hayashi, G. Eckart, G. Holzwarth, and H. Walliser, {\it{Phys.Lett.}} {\bf 147B} 
 (1984) 5;\\
 M.P. Mattis, and M.E. Peskin, {\it{Phys.Rev.}} {\bf D32} (1985) 58;\\ 
 
 \bibitem{Mattis} M.P. Mattis and M. Karliner, {\it{Phys.Rev.}} {\bf D31} (1985) 2833;\\
 G. Eckart, A. Hayashi, and G. Holzwarth, {\it{Nucl.Phys.}} {\bf A448} (1986) 732;\\
 
 \bibitem{Brodsky}S. Brodsky, J. Ellis, and M. Karliner, {\it{Phys.Lett.}} {\bf B206} (1988) 309;\\
 R. Johnson, N.W. Park, J.Schechter, V. Soni, and H. Weigel {\it{Phys.Rev.}} {\bf D42} (1990) 2998;\\
 
\bibitem{Jones} M.K. Jones et al., {\it{Phys.Rev.Lett.}} {\bf 84} (2000) 1398;\\ 
 
\bibitem{Ho96} G. Holzwarth, {\it{Z.Physik}} {\bf A356} (1996) 339;\\ 
 

\bibitem{Witten}E. Witten, {\em Nucl. Phys.} {\bf B223} (1983) 422,433;\\
 
\bibitem{Mouss}B. Moussallam, {\it Ann. Phys.(NY)} {\bf 225} (1993) 264;\\
F. Meier and H. Walliser, {\it Phys.Reports} {\bf 289} (1997) 383;\\

\bibitem{Bjorken} A.A. Anselm, {\em Phys. Lett.} {\bf B217} (1988) 169; \\
A.A. Anselm and M.G. Ryskin, {\em Phys. Lett.} {\bf B266} (1991) 482;\\
J.P. Blaizot and A. Krzywicki, {\em Phys. Rev.} {\bf D46} (1992) 246;
 {\em Phys. Rev.} {\bf D50} (1994) 442;\\
J.D. Bjorken, {\em Int. J. Mod. Phys.} {\bf  A7} (1992) 4189;\\
S.Gavin, {\em Nucl. Phys.} {\bf A590} (1995) 163c;\\


\bibitem{Bearden} I.G. Bearden et al.,  {\it Phys.Rev.} {\bf C65} (2002) 044903;\\
M.M. Aggarwal et al., {\it Phys.Rev.} {\bf C65} (2002) 054912;\\
T.K. Nayak et al., {\it Pramana} {\bf 57} (2001) 285-300;\\
T.C. Brooks et al.,  {\it Phys.Rev.} {\bf D61} (2000) 032003;\\


\bibitem{Gavin}
S. Gavin, A. Gocksch, and R.D. Pisarski, {\it Phys.Rev.Lett.} {\bf 72} (1994) 2143;\\

\bibitem{HoKl02}G. Holzwarth and J. Klomfass, {\it Phys.Rev.} {\bf D66} (2002) 045032;\\


\bibitem{DeGrand} T.A. DeGrand, {\it Phys. Rev.} {\bf D30} (1984) 2001;\\ 
J. Ellis, U. Heinz, and H. Kowalski, {\it Phys. Lett.} {\bf B233} (1989) 223;\\
J.I. Kapusta and A.M. Srivastava, {\it Phys.Rev.} {\bf D52} (1995) 2977;\\
J. Dziarmaga and M. Sadzikowski,{\it Phys.Rev.Lett.} {\bf 82} (1999) 4192;\\


\bibitem{phenix}K. Adcox et al., [PHENIX Collaboration], {\it Phys.Rev.Lett.} {\bf 88} (2002) 242301;\\ 

\bibitem{Kibble} T.W.B. Kibble, {\it J. Phys.} {\bf A9} (1976) 1387;\\
 N.H. Christ, R. Friedberg and T.D. Lee, {\it Nucl.Phys.} {\bf B202} (1982) 89;\\

\bibitem{Zurek} W.H. Zurek, {\it Nature} {\bf 317} (1985) 505;\\
G.E. Volovik, {\it Exotic Properties of Superfluid
He$^3$} (World Scientific, Singapore 1992);\\ 
M. Hindmarsh and T.W.B. Kibble, {\it Rep. Prog. Phys.} {\bf 58} (1995) 477;\\
W.H. Zurek, {\it Phys. Rept.} {\bf 276} (1996) 177;\\
P. Laguna and W.H. Zurek, {\it Phys. Rev. Lett.} {\bf 78} (1997) 2519;\\
R.H. Brandenberger, {\it Pramana} {\bf 51} (1998) 191;\\
G.J. Stephens, {\it{Phys.Rev.}} {\bf D61} (2000) 085002;\\

\bibitem{Huang} Z. Huang and X.-N. Wang, {\it Phys.Rev.} {\bf D49} (1994) 4335; \\
F. Cooper, Y. Kluger, E.Mottola, and J.P. Paz, {\it Phys.Rev.} {\bf D51} (1995) 2377; \\
M.A. Lampert, J.F. Dawson, and F. Cooper, {\it Phys.Rev.} {\bf D54} (1996) 2213; \\
J. Randrup, {\it Phys. Rev.Lett.} {\bf 77}(1996) 1226; {\it Nucl.Phys.} {\bf A616} (1997) 531; \\
T.C. Petersen and J. Randrup, {\it Phys.Rev.} {\bf C61} (2000) 024906; \\
O. Scavenius, A. Dumitru, and A.D. Jackson, {\it Phys.Rev.Lett.} {\bf 87} (2001) 182302;\\


\bibitem{Ho03} G. Holzwarth,{\it Phys.Rev.} {\bf D68} (2003) 016008; \\

\bibitem{HoKl01} G. Holzwarth and J. Klomfass,{\it Phys.Rev.} {\bf D63} (2001) 025021; \\

\bibitem{Ellis}J. Ellis and H. Kowalski, {\it Phys. Lett.} {\bf B214} (1988) 161.
\end{thebibliography}
\end{document}